\documentclass[9pt,twocolumn,twoside]{pnas-new}
\usepackage{graphics,color,graphicx,amsmath}
\usepackage{xr}
\usepackage{rotating}[counterclockwise]

\newcommand{\br}[1]{\left<#1\right>}

\newcommand{\lp}[0]{\lambda_+}
\newcommand{\lm}[0]{\lambda_-}

\templatetype{pnasresearcharticle} % Choose template 

\significancestatement{We show that firm competition, viral mutations, and scientific progress are connected through a fundamental process of innovation and obsolescence. When innovations are indexed by firm productivity, genetic mutations, or scientific contributions, they describe a space of the possible that grows and shrinks over time. We propose a reduced model of these dynamics to discover three possible worlds: one where innovation is matched by obsolescence, one of abundant diversity because innovation forever outpaces obsolescence, one of no diversity because of fast obsolescence. The boundary between the latter two can be boom and bust, where the space alternates from sparse to abundant. The first is ``creative destruction,'' where the space of the possible is a fixed size, but the size is surprisingly sensitive to innovating agent behavior. From this, we predict characteristic profiles for how many firms, viral variants, or scientific publications are near the innovative frontier. To our surprise, the examples are comparable to one another, our model aligns with data, and we find a general measure of innovativeness. Thus, we introduce a fresh and testable perspective on how systems express shared principles of innovation and obsolescence.}

\title{Idea engines: Unifying innovation \& obsolescence from markets \& genetic evolution to science}
\author[a]{Edward D.~Lee}
\author[b]{Christopher P.~Kempes}
\author[b]{Geoffrey B.~West}
\affil[a]{Complexity Science Hub, Josefst{\ae}dter Strasse 39, Vienna, Austria}
\affil[b]{Santa Fe Institute, 1399 Hyde Park Rd, Santa Fe, USA}
\dates{This manuscript was compiled on \today}
\leadauthor{Lee}

\authorcontributions{All authors collaboratively designed the research. E.D.L.~wrote the code and initially drafted the manuscript. All authors contributed to the analysis and edited the manuscript.}

\authordeclaration{The authors declare no competing interests.}
\correspondingauthor{\textsuperscript{2}To whom correspondence should be addressed. E-mail: edlee@csh.ac.at}


\begin{abstract}
Innovation and obsolescence describe dynamics of ever-churning and adapting social and biological systems, concepts that encompass field-specific formulations. We formalize the connection with a reduced model of the dynamics of the ``space of the possible'' (e.g.~technologies, mutations, theories) to which agents (e.g.~firms, organisms, scientists) couple as they grow, die, and replicate. We predict three regimes: the space is finite, ever growing, or a Schumpeterian dystopia in which obsolescence drives the system to collapse. We reveal a critical boundary at which the space of the possible fluctuates dramatically in size, displaying recurrent periods of minimal and of veritable diversity. When the space is finite, corresponding to physically realizable systems, we find surprising structure. This structure predicts a taxonomy for the density of agents near and away from the innovative frontier that we compare with distributions of firm productivity, covid diversity, and citation rates for scientific publications. Remarkably, our minimal model derived from first principles aligns with empirical examples, implying a follow-the-leader dynamic in firm cost efficiency and biological evolution, whereas scientific progress reflects consensus that waits on old ideas to go obsolete. Our theory introduces a fresh and empirically testable framework for unifying innovation and obsolescence across fields.
\end{abstract}

\begin{document}
\maketitle

\thispagestyle{firststyle}
\ifthenelse{\boolean{shortarticle}}{\ifthenelse{\boolean{singlecolumn}}{\abscontentformatted}{\abscontent}}{}

\dropcap{U}nderstanding the dynamics and structure of innovation and obsolescence has been a subject of considerable interest across many domains ranging from business, economics, and technology to evolutionary biology, medicine, and the physical sciences. 
The forces of innovation and obsolescence define classical capitalist markets, summarized in  Schumpeter's iconic term ``creative destruction,'' based on the idea that new methods of production survive by eliminating existing ones \cite{schumpeterTheoryEconomic1983}. This is echoed in Spencer's description of evolution as the ``survival of the fittest,'' and in Spielrein's ``destruction as the cause of coming into being'' for psychological development \cite{spencerPrinciplesBiology1864}. In the natural sciences, we have the more charitable adage from Newton that we build ``on the shoulders of giants.'' Each of these aphorisms implies that the new destroys or eclipses the old. A key point is that innovation of one thing often causes the obsolescence of another. A second key point is that agents, such as firms, organisms, or scientists, are themselves creating technologies, behaviors, or capacities that they could adopt from the set of the possible, while disregarding the irrelevant. 
In the substantial literature, each area has developed its own particular formalization of the problem that has masked their fundamental similarities. Furthermore, there have been few attempts to formulate the problem of innovation and obsolescence in an analytic framework that is quantitative, predictive, and testable. Indeed, one of the major obstacles for validating models of innovation and obsolescence is that they make predictions that are difficult to test empirically --- not to mention across examples as diverse as economics, genetics, and science. One of our contributions is to map proxies of innovativeness from empirical examples to model predictions of the density of agents near the innovative frontier. We demonstrate alignment with our theory using examples of firm productivity, the emergence of new clades in viral mutations, and citation rates that represent the wave of attention across the innovation front. These concrete examples surprisingly align with our theory derived from first principles, capture essential features of shared dynamics, and thus connect diverse systems within a unified theory.

\begin{figure}[t]\centering
	\includegraphics[width=\linewidth]{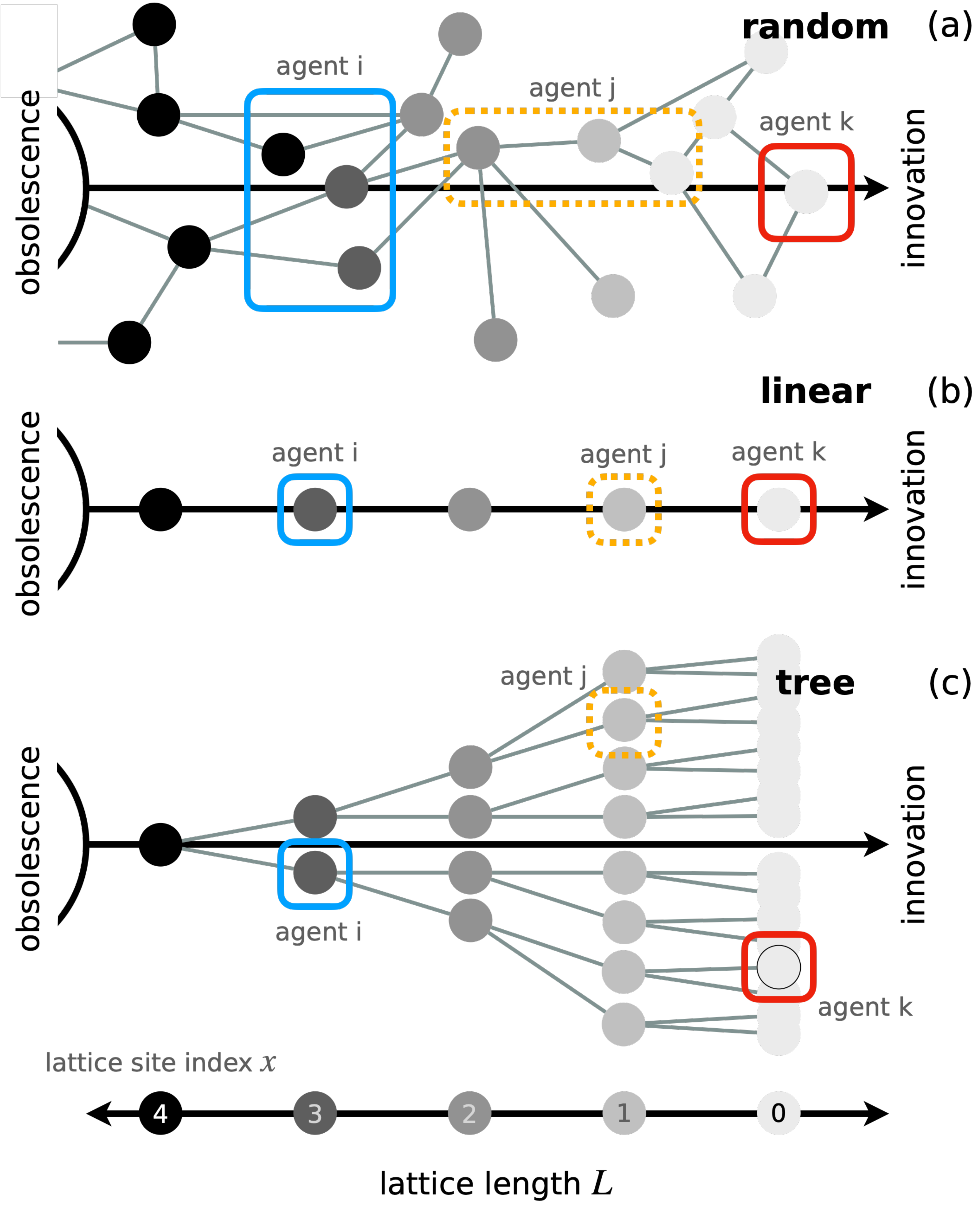}
	\caption{Model diagram on (a) a random graph structure, (b) linear lattice projection, (c) tree. (a) In the most realistic approximation, agents can span a single or multiple ideas on a random, dynamic idea lattice. We approximate this with the base case of (b) a linear lattice with length $L$ that grows along the innovation front at $x=0$ into the adjacent possible to the right and is obliterated along the obsolescence front to the left at $x=L-1$. This can be extended to other structure such as (c) trees. Lattice index $x$ is shown at bottom.}\label{gr:model}
\end{figure}

A primary distinction between fields is the relationship assumed between innovation and obsolescence. In Schumpeter's creative destruction, the relationship is one of conservation, where productive innovation comes at the expense of an existing mode of production. A realization of this is the study of economic competition as new methods of production are innovated \cite{frankeWaveTrains2001,andergassenInnovationWaves2006}. It follows that firms live on a line of productivity margins \cite{iwaiSchumpeterianDynamics1984, iwaiSchumpeterianDynamics1984a}, and obsolescence occurs endogenously either from desuetude or unsustainable profit margins \cite{aghionModelGrowth1990}. In the study of scientific progress \cite{valverdeTopologyEvolution2007,valverdePunctuatedEquilibrium2015,fortunatoScienceScience2018}, social change \cite{kolodnyGameChangingInnovations2016}, and biological evolution \cite{vendittiSpeciationBursts2008}, obsolescence is the complement of innovation: reduced citation rates imply that articles are forgotten, norms switch in a binary way from one to another \cite{amatoDynamicsNorm2018}, and extinction is a natural result of being outcompeted \cite{spencerPrinciplesBiology1864}. Other examples, however, reveal more complicated relationships between innovation and obsolescence. In markets, obsolescence may be driven by external research developments funded by government programs (e.g.~GPS, Internet, mRNA vaccines) or by products from technologically advanced neighbors. Some innovations open many more possibilities than they close \cite{hanelPhaseTransition2005}. In biological evolution, obsolescence may depend on  environmental shifts \cite{mayhewBiodiversityTracks2012}. Thus, a general theory must encompass different relationships between innovation and obsolescence of which ``creative destruction'' and complementarity are special cases.

A secondary distinction between fields is the ontological choice of what is being innovated (the space of the possible represented as an {\it idea lattice}) and who is doing the innovating (the agent). For example, a lattice site may represent a product offered or a manufacturing method used by a firm. Alternatively, the lattice could be new mutations in a population or a time-ordered list of topics that have emerged in the scientific literature. The definition of the unit of innovation has been an especially vexing problem in biology, where innovations can be genotypic, phenotypic, behavioral, or environmental \cite{erwinInsightsInnovation2004}. Such distinctions can be important. For example, resource constraints shape organism metabolism \cite{kempesPredictingMaximum2011,leeGrowthDeath2021}, or physical constraints act on the distribution of new mutations \cite{mulliganInterplayProtein2015, prielerSpo11Generates2021}. By focusing on the elementary tension between innovation and obsolescence, we integrate particular mappings and particular constraints, the details of which could be represented as mathematical relations between a general set of dynamical parameters.

\section*{A generalized model from first principles}
We picture innovation and obsolescence to take place in a space of possibilities, an idea lattice, in which agents live and that is itself constantly churning as shown in Fig.~\ref{gr:model}a. Here, each vertex $x$ represents an ``idea'' in which an agent is invested. Ideas share an edge with related ideas, denoting either material similarity, shared inputs and skills, or common ancestry \cite{teeceUnderstandingCorporate1994}. Clustering of similar ideas allows us to compress related items into a single site, which leads to the linear lattice approximation in Fig.~\ref{gr:model}b. This simplification is equivalent to choosing a scale of analysis, where agents can be neatly assigned to sets such as by clustering them with industry sector codes or with function for genetic mutations \cite{hidalgoProductSpace2007,neffkeHowRegions2011}. Then, we define an obsolescence front at which ideas go defunct (vertices removed) and an innovation front at which agents drive ahead into the ``adjacent possible''  (vertices added) \cite{kauffmanInvestigations2000}. 
%An example of this is from empirical studies of export diversity, where has been shown that the largest eigenvalues of national export diversification constitutes a typical unidimensional direction in increasing manufacturing concentration \cite{hausmannWhatYou2007,mcnerneyBridgingShortterm2021}. 

\begin{figure}[t]\centering
	\includegraphics[width=.9\linewidth]{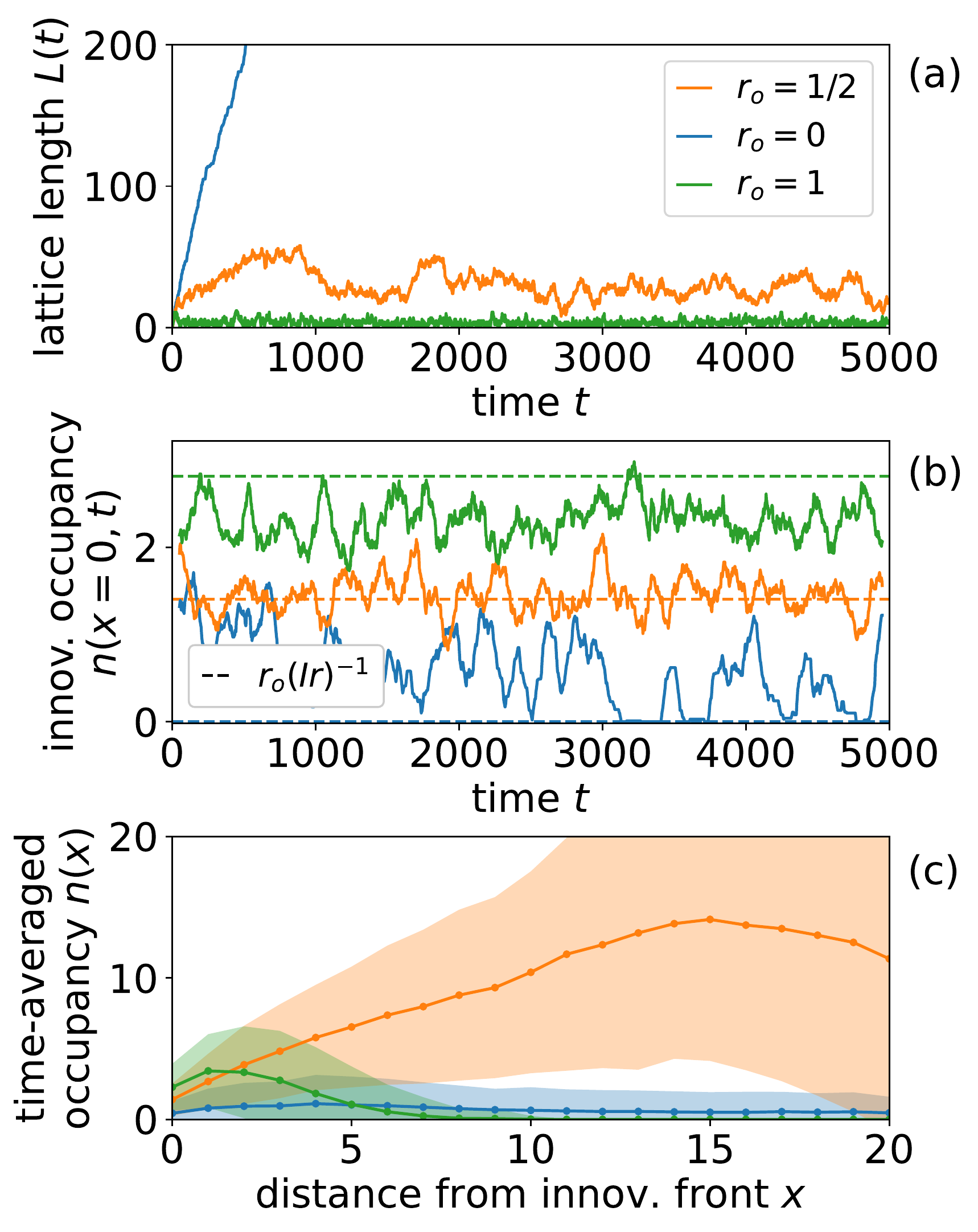}
	\caption{Static, runaway, and collapsed dynamics in automaton simulation measured by (a) lattice length, (b) occupancy number at innovation front, and (c) pseudogap shape averaged over time. (b) Dashed lines indicate steady-state conditions, which obsolescence rate $r_o=1/2$ satisfies. (c) Error bars, shaded areas, show standard deviation over $10^3$ time steps. Colors indicate the same three conditions across all panels with orange for stasis, blue for runaway, and green for collapsed. Parameters growth rate $G=4$, replication rate $r=0.395$, death rate $r_d=0.4$, innovativeness $I=0.9$.}\label{gr:sim example}
\end{figure}

We consider simple agent dynamics that represent their mean tendencies to grow, die, and replicate in a way that summarizes more detailed processes. New agents such as firms, mutants, or publications enter the system with a rate $G$ distributed uniformly amongst the number of lattice sites $L(t)$ at time $t$. Agents may replicate into the adjacent innovative site with rate $r$ --- for example, by copying more innovative ideas --- and leave the system with death rate $r_d$. Our treatment considers these as two basic independent parameters whose interrelationship maps to a wide range of scenarios.

At the innovation front, agents encounter the additional difficulty of inventing what is possible before occupying it. In the case where agents innovate independently, the rate of successful innovation is proportional to how often they seek to expand, $r$, their innovativeness, $I$, and the number of agents at the innovation front, $n(x=0,t)$, thereby summarizing the complex process of discovery in terms of a mean rate \cite{younInventionCombinatorial2015,iacopiniInteractingDiscovery2020} (see Appendix~\ref{si sec:model derivation}). 

Finally, we incorporate obsolescence by assuming that the oldest idea goes obsolete with a rate $r_o$. Consequently, the rate of change of the lattice length $\dot L(t)$ is the difference between the rates of innovation and obsolescence,
\begin{align}
	\dot{L}(t) &= rIn(0,t) - r_o. \label{eq:L(t)}
\end{align}
A minimum allowed length of $L=1$ corresponds to when the two fronts coalesce. When the lattice length is stable $\dot L(t)=0$, the system looks like an ``innovation train'' moving into the innovation frontier with innovativeness directly related to obsolescence with 
\begin{align}
	n(0,t) &= r_o (r I)^{-1}. \label{eq:stat n0}
\end{align}
To simplify the mathematical treatment, we imagine ``sitting'' on the train and fixing the coordinate system such that the innovation front is at the origin $x=0$ with its movement represented by the train tracks moving past us to the right (this reverses the coordinate system from left to right from what is depicted in Fig.~\ref{gr:model}). Putting these together, the rate of increase in the number of agents $\dot{n}(x,t)$ at lattice site $x$ at time $t$ is
\begin{align}
\begin{aligned}
	\dot n(x, t) &= \frac{G}{L(t)} + r n(x+1, t) -r_d n(x,t) - \\
	&\qquad r I n(0,t) [n(x,t) - n(x-1,t)].
\end{aligned}\label{eq:1d lattice}
\end{align}
The first term is the rate at which new firms enter the system, the second the rate at which they expand by mimetic innovation \cite{dawkinsSelfishGene2016}, the third the rate at which they leave the system, and the last term the effective shift from the motion of the innovation front. We solve Eqs~\ref{eq:L(t)} and \ref{eq:1d lattice} using both analytic approximation and numerical calculation including simulations that are further detailed in Appendix~\ref{si sec:num calc}. These results summarize the mean-field dynamics of a simple system of agents growing, dying, and innovating in a one-dimensional space. Below, we show how these equations can be straightforwardly generalized to include more complex dynamics such as cooperative innovation, higher-dimensional graphs, and inverted obsolescence-driven innovation.

\section*{Creative destruction, runaway innovation, \& collapse}
Eq~\ref{eq:L(t)} predicts three regimes of idea graph dynamics: i) innovation and obsolescence are roughly balanced, leading to a typical size of the idea space; ii) innovation outpaces obsolescence and the system grows indefinitely, providing an unbounded ``marketplace'' for exploitation; iii) obsolescence outpaces innovation and the system collapses to only a few, transient ideas. In order to demonstrate the three regimes, we provide three examples of a stochastic automaton simulation following the dynamics specified in  Eqs~\ref{eq:L(t)} and \ref{eq:1d lattice}, and illustrated in Fig.~\ref{gr:sim example} (Appendix~\ref{si sec:num calc}). When the rate of obsolescence $r_o$ is sufficiently small (blue line), we are in the regime of runaway innovation. As we increase $r_o$, we pass through a regime of steady lattice length (orange line) to a small lattice that repeatedly collapses to its minimum size $L=1$ (green line). As a point of departure for analysis, we start with a stationary configuration in which each innovation extinguishes one old idea, the assumption underlying Schumpeter's original formulation of creative destruction \cite{schumpeterTheoryEconomic1983}.

The stationary condition is a fundamentally collective property. In Eq~\ref{eq:stat n0}, global stability means that the number of agents on the leading edge is proportional to the rate at which ideas go obsolete, i.e.~faster obsolescence means more highly innovative agents to sustain rapid progress. Alternatively, faster innovation leads to a drop in the number of innovative agents because fewer agents keep innovations apace. Under stochastic variation, this observation means that a temporary increase in the number of innovative agents $n(0,t)$ above steady state drives the innovation front ahead quickly, consequently reducing the number of innovative agents. This reduced number then slows the innovation front down and allows for agents to flow in from the existing lattice, resulting in oscillations around steady state occupancy (Fig.~\ref{gr:sim example}b). Thus, the age of an idea near the innovation front is naturally related to the number of agents because newer sites have fewer agents.\footnote{This result agrees with the intuition, for example, that firms expand into ``adjacent markets'' because there is less competition \cite{barroEconomicGrowth2003}.}

Following the logic that older ideas will have accumulated more agents, we expect the typical number of agents to be minimal near the innovation front and to increase as we move towards older ideas. This describes what in physics is referred to as a ``pseudogap'', namely a drop in the density of excited states, around $x=0$. If this is assumed to be approximately linear at stationarity, i.e.~$\Delta n(x=0) \approx$ constant, then
$n(1) - n(0) \approx n(0) - n(-1) $. Since, by definition, the site at $x=-1$ is unoccupied, i.e.~$n(-1) = 0$, this gives $n(1) \approx 2 n(0) $. Setting $x = 0$ in Eq~\ref{eq:1d lattice} we can then solve for lattice length at stationarity:
\begin{align}
	L &= GIrr_o^{-1} \left[ r_o+r_d-2r \right]^{-1}.\label{eq:L}
\end{align}
While Eq~\ref{eq:L} is only an approximation to the exact nontrivial solution, which manifests surprisingly complex variation (see Appendix~\ref{si sec:model derivation} and Fig.~\ref{si gr:2nd ode phase space}), it does provide a rather revealing starting point for capturing the essential characteristics of the resulting idea space. 

\begin{figure}\centering
	\includegraphics[height=\linewidth]{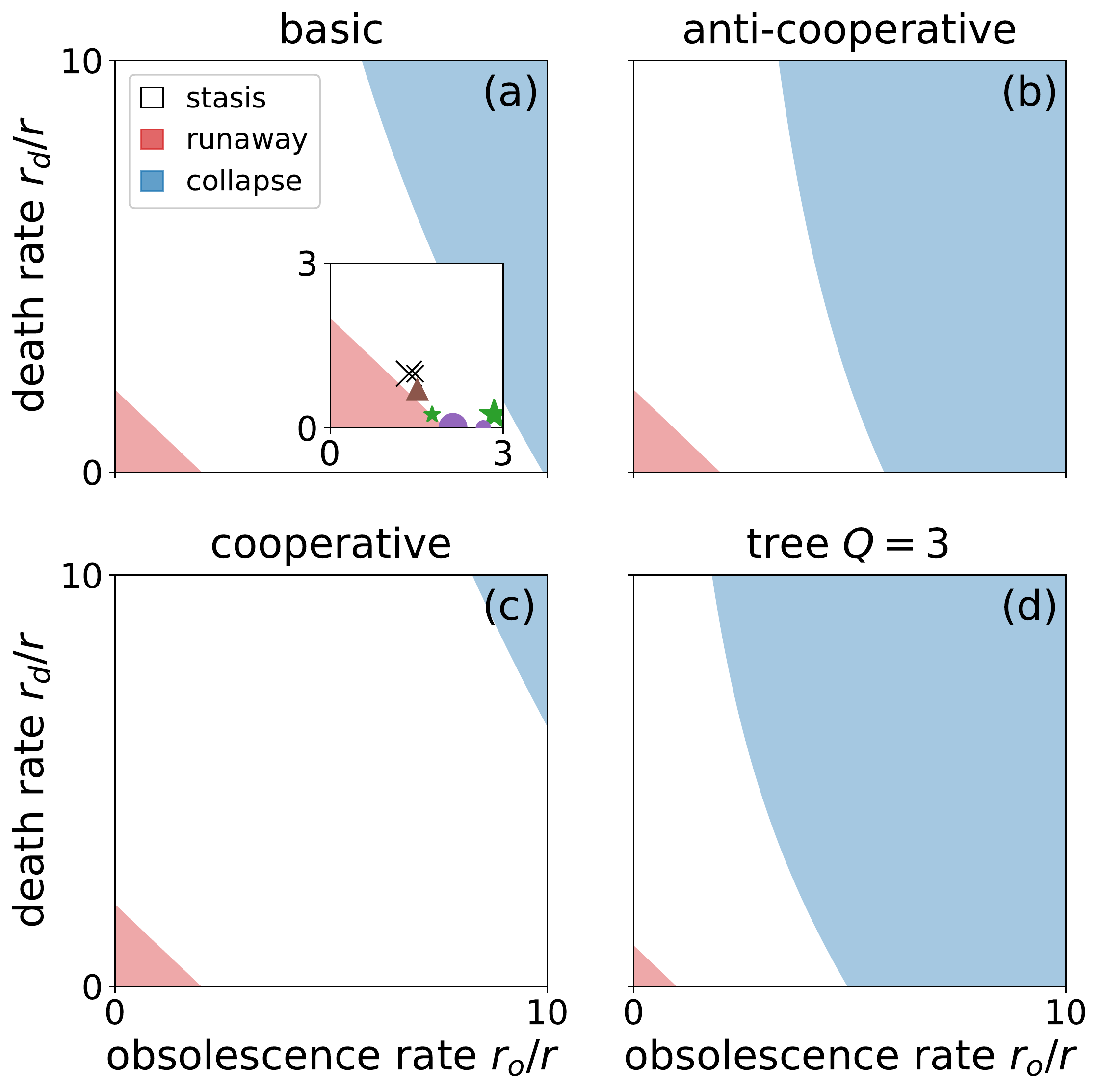}
	\caption{Model phase space. (a) Basic model predicts regimes in which the size of the space of the possible, or the lattice length, stabilizes (white), grows indefinitely (red), or collapses (blue). Parameter values rescaled growth rate $G/r=1$, innovativeness $I=1$. (inset) Empirical examples located in the phase space include metal stamping firms from Fig.~\ref{gr:pseudogap examples}d (brown triangle), Indian firms from Fig.~\ref{gr:pseudogap examples}g (small/large green stars for small/large firms), patent citation curves from Fig.~\ref{gr:pseudogap examples}f (least/most as small/large purple circles), and {\it Physical Review B} citation curves from Fig.~\ref{gr:pseudogap examples}j (few/many citations as small/large black X's). Model extensions to (b) anti-cooperative innovation, $\alpha=1/2$, (c) cooperative innovation, $\alpha=3/2$, and (d) tree graph with branching ratio $Q=3$.}\label{gr:phase space}
\end{figure}

Two pivotal points are indicated by Eq~\ref{eq:L}: (i) where obsolescence is perpetually outpaced by innovation and $L\rightarrow\infty$; and (ii) where obsolescence outpaces innovation and $L\rightarrow 1$ signaling the collapse of the system. Formally, the first can occur when $G$, $I$, or $r$ become infinite, or when $r_o=0$, none of which is realistic because they require infinitely fast rates or perpetual suppression of obsolescence. On the other hand, the singularity at $r_o+r_d-2r = 0$ requires only balancing $r_o/r$ and $r_d/r$ such that
\begin{align}
	r_o/r \leq 2 - r_d/r. \label{eq:divergence inequality}
\end{align}
We delineate this region in red in Fig.~\ref{gr:phase space}. Eq~\ref{eq:divergence inequality} indicates that the typical number of times an agent reproduces before it dies is a crucial order parameter and that agent-level properties can drive unbounded growth of the idea lattice, leading to runaway innovation (the first pivotal point).

The second pivotal point is reached when the innovation front number falls below a self-sustaining threshold, $n(0,t) < (r/r_o ) I $. From Eq~\ref{eq:L(t)}, growth becomes negative, $\dot L(t) < 0$, driving the system to collapse to its minimum length ${L\sim1}$. From Eq~\ref{eq:L}, we can solve for the corresponding cutoff obsolescence rate as
\begin{align}
	r_o/r &\sim 1 -r_d/2r + \sqrt{(1-r_d/2r)^2 + GI/r}.\label{eq:obs rate}
\end{align}
In the limit $GI/r\rightarrow0$, the boundary between collapse and growth shrinks to a line and eliminates the region of stability (white region in Fig.~\ref{gr:phase space}a). Consequently, we expect to find a sharp transition between collapse and growth, and the system can display long time  scales and large fluctuations \cite{thurnerSchumpeterianEconomic2010} (Appendix~\ref{si sec:fluctuations}). Thus, Eqs~\ref{eq:divergence inequality} and \ref{eq:obs rate} elucidate the three regimes of lattice dynamics: a balanced, a runaway, and a collapsed regime.

\section*{Extensions}
This picture holds for several key generalizations such as cooperative innovation, higher-dimensional graphs, and inverted obsolescence-driven innovation, a reversed picture where obsolescence furthers system progress, as we clarify below: 

(i) Cooperative innovation implies that the front velocity scales nonlinearly with the number of agents as $r I n(0)^\alpha$. 
%Here superlinearity, $\alpha>1$, signals cooperation, whereas sublinearity, $\alpha<1$, implies competition. 
The default value of $\alpha=1$, as in Eq~\ref{eq:1d lattice}, corresponds to agents innovating independently of one another. Here superlinearity, $\alpha>1$, signals cooperation, whereas sublinearity, $\alpha<1$, implies competition. From Eq~\ref{eq:stat n0}, these differences can be mapped back to Eq~\ref{eq:1d lattice} by the transformation $I^\alpha\rightarrow I$ and $(r_o/r)^\alpha\rightarrow r_o/r$. 

(ii) For tree graphs, shown in Fig.~\ref{gr:model}c, each sequential site branches into $Q-1$ additional branches, one of which must be chosen by a new agent. If branches are equally likely, the replication term in Eq~\ref{eq:1d lattice} acquires an additional factor $(Q-1)^{-1}$ (see Eq~\ref{si eq:bethe lattice}), such that the number of agents systematically decreases towards the innovation front. This argument makes clear the importance of the relative dimensions of agent replication and the idea space. When next-generation agents do not fill all of the available space, then agents necessarily occupy a small fraction of the idea space. Nevertheless, the dimensional depletion effect does not fundamentally alter the dynamics. If we rescale $r\rightarrow (Q-1)r$ and $I\rightarrow I/(Q-1)$, then we again recover Eq~\ref{eq:1d lattice}. As argued in Appendix~\ref{si sec:extensions}, higher Euclidean dimensions can also be approximated by the linear model. Thus, important classes of dynamical or structural generalizations do not appreciably alter the basic model.

\begin{figure*}[t]\centering
	\includegraphics[width=.85\linewidth]{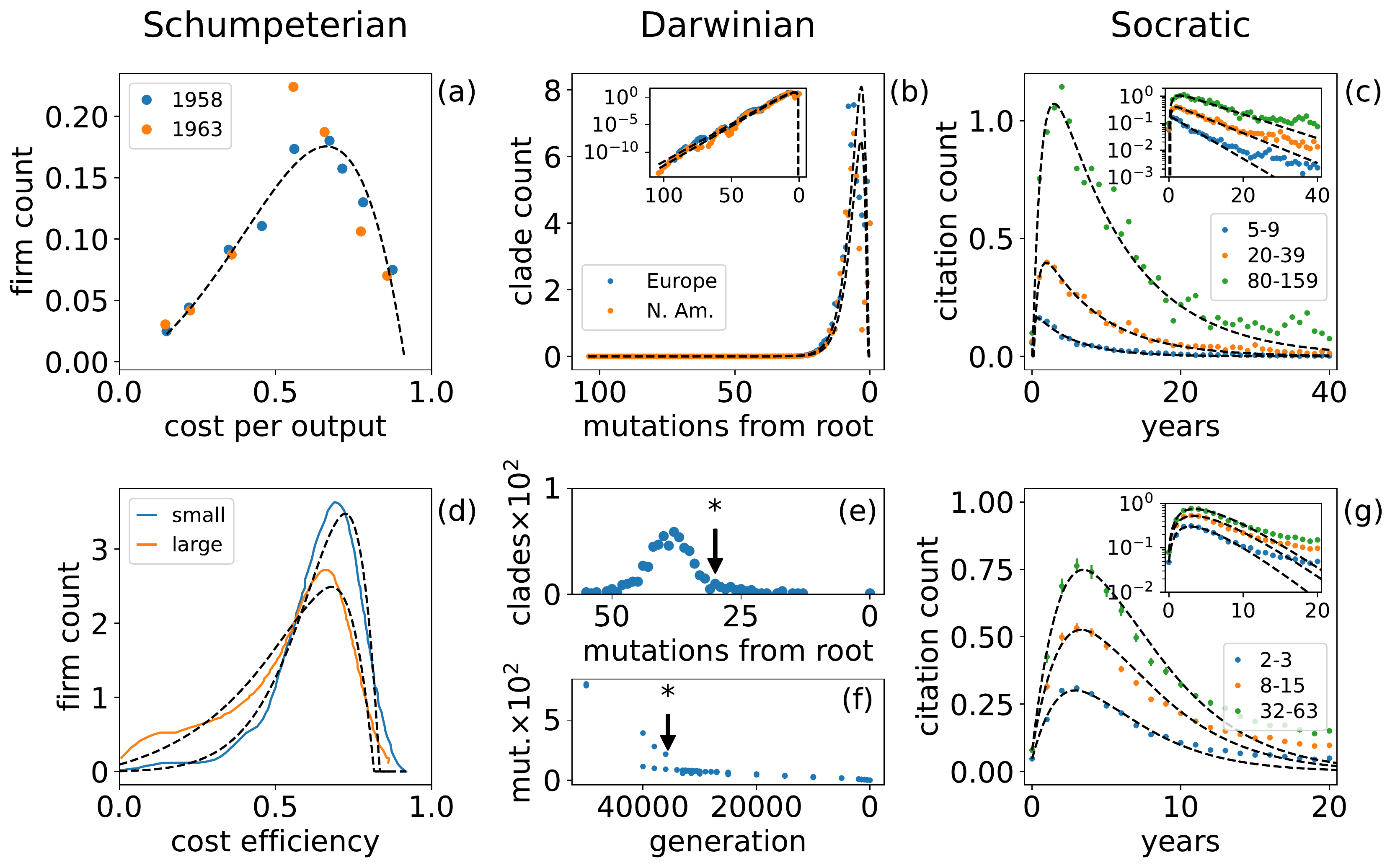}
	\caption{Empirical vs.~model predictions for innovativeness pseudogaps in (a,\,d) firms, (b) genetic mutations, and (c,\,g) scientific citations. Data is in color and model fits are dashed black lines. Histogram for (a) U.S.~metal stamping industry with model fit (black) to the year 1958 \cite{iwaiSchumpeterianDynamics1984} and (d) cross-industry Indian firms grouped by size \cite{jangiliImpactSize2019}. (b) SARS-CoV-2 diversity by number of base pair mutations from first detected strain hCoV-19/Wuhan/Hu-1/2019 normalized by branching rate \cite{hadfieldNextstrainRealtime2018, sagulenkoTreeTimeMaximumlikelihood2018}. Model overpredicts occupancy number near the origin, implying variants were missed by limited surveillance at the beginning of the pandemic \cite{mallapatyWhereDid2022}. (c) Normalized citation rate for {\it Physical Review B} articles published in 1980 and separated by total citations as of 2020. (g) Normalized citation rates for patent applications in electrical and electronic technologies filed in 1990 and granted before 2022 separated by total citations within 20 years of filing \cite{hallNBERPatent2002}. Sharp Einsteinian innovation events change the very parameters of innovation such as with (e) the emergence of strains in the Delta clade and (f) fast mutator strain in long-term {\it E. Coli} experiments where starred \cite{blountGenomicAnalysis2012}. Further details in Methods.}\label{gr:pseudogap examples}
\end{figure*}

(iii) Obsolescence-driven innovation is the antithesis of forward-looking innovative dynamics. By reversing the direction of the $x$-axis in Fig.~\ref{gr:model} and setting  the innovation front at the origin, agents now replicate towards obsolescence. For stable pseudogaps, the introduction of a new idea propels every agent towards the innovative front, making the entire system more innovative. As a result, innovation is driven at a rate proportional to the number of agents on the verge of extinction, which is still given by Eq~\ref{eq:stat n0}. Agents occupying newer ideas tend to beget agents on older ideas, which then drive themselves to extinction by eventually increasing occupancy at the obsolescence front. This is evocative of Kuhnian scientific progress, where invalidation of old ideas permits new ideas, new ideas stimulate revision of existing topics, and the system as a whole progresses \cite{fosterTraditionInnovation2015}. %In the end, a seemingly different set of dynamics, obsolescence-driven innovation, is a reinterpretation of the same underlying equation.
%Interestingly, agents at the innovation front disappear into the adjacent possible as if there is a threshold beyond which innovative ideas are no longer recognized. Indeed, this agrees with the notion that true novelty is difficult to recognize and reflects the tendency of experts to undervalue novelty \cite{boudreauLookingLooking2016}. 

\section*{Innovation distributions in model \& data}
%The fact that Eq~\ref{eq:1d lattice} can be easily transformed to account for structural, dimensional, and reversed extensions of the model implies that its resulting pseudogap form predicts a generalizable taxonomy. We find that the taxonomy of creative destruction is highly constrained, consisting of an exponential rise to a skewed peak with a decaying tail or a sublinear rise to a flattened occupancy function characterized by agents piling up near the innovation front (see Appendix Fig.~\ref{si gr:pseudogap shapes}). The two forms constitute types of the innovative adjacent possible, which are complemented at the other end by types of the adjacent obsolescent.\footnote{Other than the sublinear and superlinear pseudogaps, the pseudogap effectively disappears as lattice length diverges.} We plot several empirical examples in Fig.~\ref{gr:pseudogap examples} of innovation distributions in markets (Schumpeterian), genetic evolution (Darwinian), and science (Socratic) in terms of occupancy number vs.~proxies for their distance from the innovation front. 

Our model predicts several characteristic forms for the density of agents near the adjacent possible, or the shape of the pseudogap $n(x)$, that holds across structural, dimensional, and reversed extensions of our model (see Appendix Fig.~\ref{si gr:pseudogap shapes}). Does this predicted taxonomy align with empirical examples?

We compare with several examples in Figure~\ref{gr:pseudogap examples} including from firms, viral genetic evolution, and technology and science, putting together examples that have so far been considered independently in the literature. In each of the panels and with the dashed black lines, we show that the functional form of the model --- a characteristic exponential rise to a skewed peak with a decaying tail --- aligns surprisingly well with the data. As we discuss in further detail below, we must determine the appropriate axis along which to map the idea lattice coordinate. For firms, we use histograms along a economic proxy for innovativeness, although the outcome depends on whether one aggregates industries or not (panels a and d). In genetic evolution, a natural measure of innovativeness is in terms of the number of base-pair mutations that distinguish a particular viral strain. Then, plotting the number of unique strains per number of mutations leads to a histogram relative to the innovative frontier, which is the most genetically distant strain from origin (panel b). Finally, in science and technology, we do not explicitly map the idea lattice. Instead, we follow in the footsteps of previous work to consider papers and patents as innovative combinations of ideas \cite{fosterTraditionInnovation2015,younInventionCombinatorial2015}. This implies that papers are typically somewhere on the frontier, and so the citation rate serves as a proxy for the density of agents near the frontier (panels c and g). We go through each example in more detail below to explain how we unify these examples within the context of our model.

Economics provides the classic example of {\it Schumpeterian} innovation measured in terms of cost efficiency, or the ability of a firm to extract profit from a fixed amount of investment \cite{farrellMeasurementProductive1957,aghionModelGrowth1990,iwaiSchumpeterianDynamics1984, iwaiSchumpeterianDynamics1984a}. This imposes a seemingly natural one-dimensional axis for ordering firms, where more innovative firms progressively decrease cost per unit output. Fig.~\ref{gr:pseudogap examples}b shows an example of the distribution of labor costs per value added for the US metal stamp industry in 1958 and 1963 from Iwai's classic work \cite{iwaiSchumpeterianDynamics1984, iwaiSchumpeterianDynamics1984a}. The fewest number of firms are the most and the least cost-efficient, although the distribution is skewed to the right because many firms with higher costs survive. This characteristic form also appears in recent distributions of Indian firms in Fig.~\ref{gr:pseudogap examples}g, but when plotted against cost efficiency, which is essentially the reversed x-axis \cite{jangiliImpactSize2019}. In this case, larger firms as measured by total assets are typically less cost efficient than smaller firms. While this would be taken as evidence of poor innovation in the canonical sense, fit of the density curves with our model indicates the opposite, that inefficient firms are more innovative (see Methods for details about fitting procedure). This is consistent with the narrative that disruptive technologies are inefficient to develop and only later are produced efficiently. While this seems to be at odds with the first example, Iwai's histograms are specific to a single industry, whereas Indian firms are aggregated across industries. More generally, innovativeness is a multi-faceted concept with many definitions in the economics literature \cite{coadAppropriateBusiness2011}. Our dynamical formulation provides a derivation of the shape of economic distributions as a function of innovative distance and thus a way of resolving discrepancies by inferring the innovative frontier from data.

As an example of biological, or {\it Darwinian}, innovation, we consider the tree of SARS-CoV-2 clades measured from the GISAID repository of sequences \cite{hadfieldNextstrainRealtime2018, sagulenkoTreeTimeMaximumlikelihood2018}. The set of possible innovations, measured by base pair mutations, is a phylogenetic tree with a branching ratio $Q\approx 2.3$ per unit phylogenetic branch length (see Methods). %Since sequencing coverage was not uniform throughout time, we do not use the date at which SARS-CoV-2 sequences were entered into the database. 
We take the number of base pair mutations from the first known strain hCoV-19/Wuhan/Hu-1/2019 to order mutants in innovative order. Then, our model presents a recursive relation (Eq~\ref{si eq:iter sol1}) that determines how the number of strains with $k-2$ and $k-1$ mutations determine the number of strains with $k$ mutations. As shown in Fig.~\ref{gr:pseudogap examples}e, the resulting occupancy plot again shows the same aforementioned characteristic form, which our model fits closely over nearly the entire period of observation.

As confirmation of what our model leaves out, we fail to capture the sharp, temporary increases when variants such as Delta emerge, indicating an important deviation from model predictions: these punctuated, unpredictable, and singular innovations violate our model assumptions --- {\it Einsteinian} innovations represent a discontinuous shift in parameters \cite{kolodnyGameChangingInnovations2016}. We highlight an example in Fig.~\ref{gr:pseudogap examples}h, where only the number of Delta clades are shown as a function of mutations from root. A similar Einsteinian innovation is revealed in long-term evolution experiments, where the sudden emergence of a mutator strain increases the  mutation rate  in an {\it E.~Coli} population, indicated by the discontinuity in accumulated mutations in Fig.~\ref{gr:pseudogap examples}i \cite{blountGenomicAnalysis2012}. The punctuated changes in innovation rates reveal a meta-dynamic in the space of parameters notably specific to genetic innovation.

Science and technology, in contrast, builds on an edifice. New technological and scientific ideas must be tethered to the past and are often judged by their consistency with established knowledge, theory, and pedagogy  \cite{kuhnEssentialTension2000}. It is only when existing frameworks have been proved insufficient that a new idea can flourish. This suggests a reversed dynamics, where innovation is driven by the obsolescence of old ideas. 

Scientific and patent citation rates support this picture as we show in Fig.~\ref{gr:pseudogap examples} panels c and g. Citation rates peak to maximal prominence quickly then slowly fall out of favor with age, a horizontally mirrored version of the previous examples. In our formulation, we take the citation profile to be proportional to the occupancy function which measures the wave of agents moving across a graph of sequential papers that mark the progression to new ideas \cite{fosterTraditionInnovation2015}. Since papers contribute different levels of innovation attributed to some intrinsic fitness \cite{valverdeTopologyEvolution2007,wangQuantifyingLongTerm2013,highamFameObsolescence2017}, we bin them based on citation count. Furthermore, we account for citation inflation (see Methods). We show yearly variation in normalized citations received by scientific papers in {\it Physical Review B} published in 1980 in Fig.~\ref{gr:pseudogap examples}c and for patent applications filed in 1990 in electrical and electronics in Fig.~\ref{gr:pseudogap examples}g. To test the predictive power of our model, we fit to only the first quarter of the duration shown, or the first decade after publication for {\it PRB} and the first five years after publication for patents. Our model fits remarkably well the beginning of the citation curves and captures the generic shape of the tails, but tends to under-predict their longer-term behavior. Possible explanations for the deviations are unconsidered effects such as bimodal memory \cite{candiaUniversalDecay2019} and debated ``runaway'' events \cite{golosovskyRunawayEvents2012,paroloAttentionDecay2015,highamFameObsolescence2017}. In addition, citation rates are biased in more recent years because citations from the newest papers and patents are missing.\footnote{Accounting for such effects could to drive down the tail by increasing the normalization factor (Methods).} This bias is stronger for patents because we do not have information about applications that are pending review, a process that typically takes several years. Yet remarkably, innovation and obsolescence dynamics lead to a first-principles, predictive explanation for citation curves. 

These examples demonstrate how the dynamics of innovation and obsolescence align with the distributions of social and biological agents from the innovative to the obsolete.

\section*{Distance to runaway innovation}
A natural way to compare the systems is to measure their distance $\Delta$ from the boundary of runaway innovation as an indicator of innovativeness. We define the distance in terms of the change in the rescaled parameters $r_o/r$ and $r_d/r$ required to reach the threshold defined in Eq~\ref{eq:divergence inequality} or visually the distance from the red region in the examples plotted in the inset of Fig.~\ref{gr:phase space}a. It is only determined by the aforementioned rescaled rates, allowing us to compare systems that differ along other dynamical parameters, growth and innovation rates.

\begin{figure}\centering
\includegraphics[width=.9\linewidth]{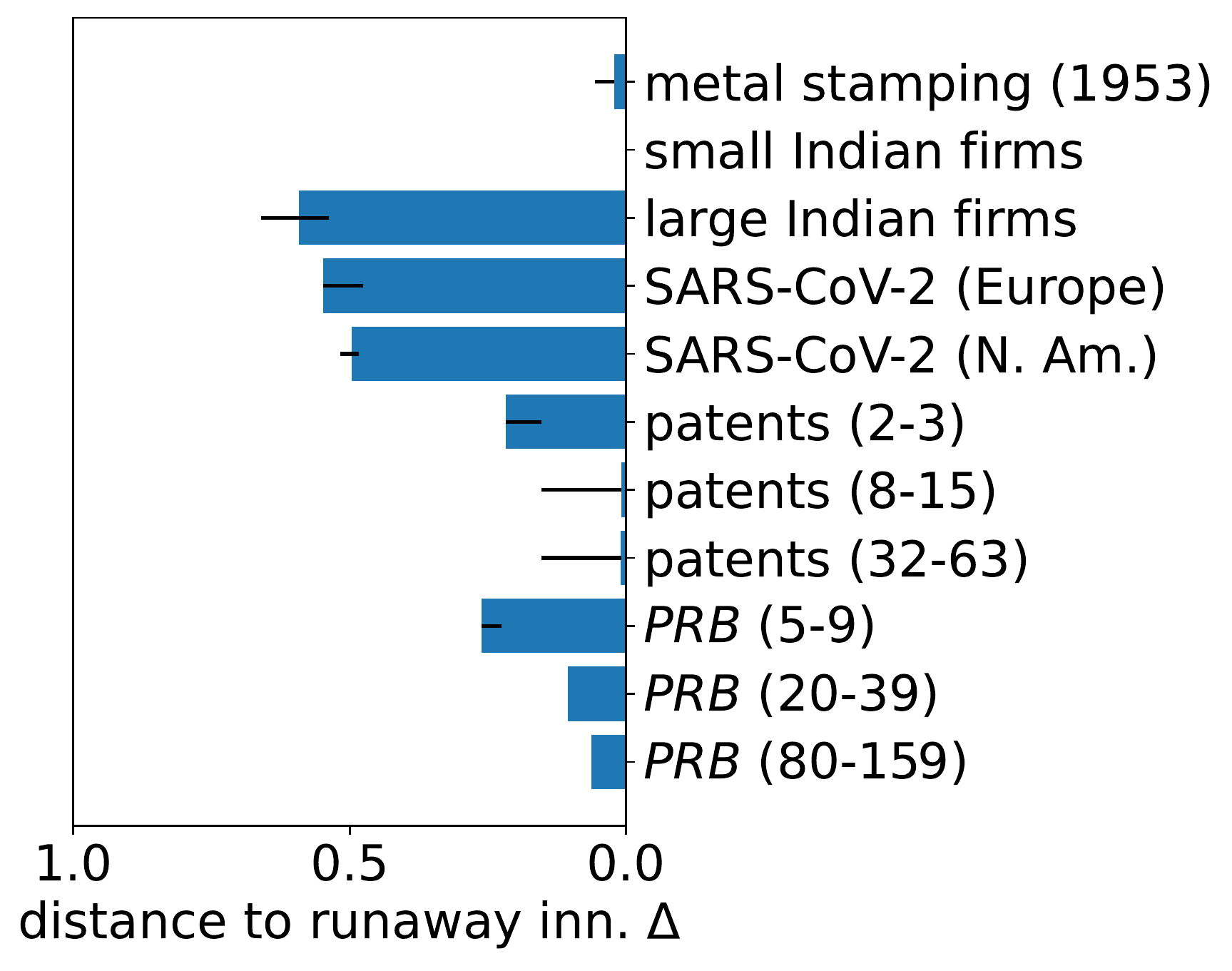}
\caption{Distance to runaway innovation for best fit model. Numbers following patents and {\it Physical Review B} ({\it PRB}) indicate the range of citation counts as in Fig.~\ref{gr:pseudogap examples}. Error bars represent variability in the fit as the minimum and maximum of the five parameter sets with smallest fit residuals.}\label{gr:div}
\end{figure}

Reassuringly, we find similar distances between the dynamics of SARS-CoV-2 evolution in Europe and North America as we show in Fig.~\ref{gr:div}. This aligns with our expectations that the overall dynamics of viral evolution were alike. In contrast, small and large non-financial Indian firms differ, and this difference extends far beyond the variation in the parameters measured across the ten closest fits. The minuscule $\Delta$ for small Indian firms suggests that they, unlike large ones, lie at the boundary of creative destruction, a critical boundary at which we predict enhanced fluctuations in cost efficiency and correspondingly of population size.

When comparing patents and {\it PRB}, we find that documents with more citations tend to lie closer to runaway innovation. This difference is especially pronounced for patents when comparing those with fewer than eight citations versus those with at least eight (see Fig.~\ref{si gr:cites} for all citation classes). A similar pattern disparity appears for {\it PRB}, although the distance decays slower with citation number. This observation suggests patents with fewer citations are on the whole more innovative than of scientific citations. As has been noted elsewhere, a scientific citation is not just an indicator of innovativeness because they can refer to established pedagogy, corrections, substantiating evidence, etc.~in contrast with how patents tend to cite patents \cite{trajtenbergPennyYour1990,aksnesCitationsCitation2019}. %Beyond universal features in citations rates \cite{wangQuantifyingLongTerm2013,highamUnravelingDynamics2017,candiaUniversalDecay2019}, we find that variation in the shape of the curves could indicate innovativeness.
Our findings highlight variation in innovativeness from systematically comparing seemingly unrelated systems as a demonstration of our generalized dynamical formulation.

\section*{Discussion}
Constituents of social and biological systems constantly undergo turnover, expanding, exploiting, or reducing the space of realized capabilities in a sequence of innovations that eventually renders the innovator obsolete. In any particular system, the exact details of this process may be modeled by competition \cite{aghionModelGrowth1990}, innovative risks \cite{rzhetskyChoosingExperiments2015}, resource constraints \cite{bloomAreIdeas2020}, limited attention \cite{golosovskyGrowingComplex2017,paroloAttentionDecay2015}, and strategy \cite{schererLinkGross2001}, amongst other factors. Furthermore, we know that the process of innovation displays combinatorial dynamics \cite{triaDynamicsCorrelated2015,younInventionCombinatorial2015,iacopiniInteractingDiscovery2020}, depends on the topology and dimension of the adjacent possible \cite{kauffmanNKModel1989,fosterTraditionInnovation2015}, and is influenced by agent interaction \cite{iacopiniInteractingDiscovery2020,ubaldiEmergenceEvolution2021}. While important, the plethora of factors obscures the fact that the same effective dynamics couple agents with the lattice on which they live (Fig.~\ref{gr:model}). The power of our generalization originates from incorporating these processes into mean rates that highlight three fundamental innovative regimes (Fig.~\ref{gr:phase space}) common across dynamical, dimensional, and structural extensions of the basic model.
% In the flavor of our work, are generalizable physics models for innovation and creative destruction \cite{thurnerSchumpeterianEconomic2010,silverbergPercolationModel2005}

When we focus on the region of parameter space corresponding to creative destruction --- where each innovative advance is matched by obsolescence --- we find a surprisingly rich taxonomy for agent density around the innovative frontier (Fig.~\ref{si gr:2nd ode phase space}). In the region of parameter space where replication rate dominates over death, or $r_d/r<1$, we find a skewed density characterized by a peak opposite a decaying tail that extends to the innovation front. Remarkably, the shape aligns with empirical examples on firm productivity, genetic evolution, and scientific citation (Fig.~\ref{gr:pseudogap examples}). Our provides one explanation for why we would expect to find empirical similarity. Thus, our framework provides a way of unifying phenomena that have so far been studied separately.

As one consequence, we can locate the examples within the same parameter space and compare them systematically. For instance, our fit parameters indicate that the systems are distinct in terms of the ratio of death to replication rate $r_d/r$, where a value close to unity can correspond to a dynamic if agents hopping from lattice site to lattice site. We measure for {\it PRB} citations $r_d/r>0.98$ and for metal stamping firms $r_d/r=0.7$. For citations, this would mean that agents progressively become less innovative over time, whereas for firms it would mean the opposite because the same firms are becoming more cost efficient over time. In contrast, the other examples indicate many replication events per site such as for Indian firms ($r_d/r\approx1/4$), SARS-CoV-2 ($r_d/r<1/20$), and patents ($r_d/r<10^{-2}$). This suggests that these systems are dominated by newcomers building on past innovations. As a second point of comparison, we define a distance to runaway innovation (Fig.~\ref{gr:div}). We find that SARS-CoV-2 mutations were similarly innovative in both North America and Europe. In contrast, we find that small, non-financial Indian firms are more innovative than large ones and that patents are typically more so than {\it PRB} articles with a comparable number of citations. In the context of debates about unleashing innovative forces in science and the economy, such metrics are a first step in deciding how the systems might be modified to promote certain collective outcomes. Finally, we find all the systems are relatively close to the boundary of runaway innovation. Whether or not this feature is representative of the ensemble of natural systems (e.g.~because evolutionary pressures drive systems towards diversity) or if researchers tend to study vibrant systems are amongst the many questions that our analysis raises.

%These observations present new questions and predictions to be tested on more fine-grained data on agent behavior.

Importantly, our theory fails to capture sudden changes that are not included such as in genetic mutation rates. Such effects could reflect unpredictable Einsteinian revolutions that modify the parameters of innovation, obsolescence, competitive effects, or atypical scientific papers with unusual citation trajectories. The deviations present additional questions to investigate as extensions of our basic model. %Nevertheless, we find it surprising that our minimal theory and correspondingly constrained functional form accurately describe empirical examples.

A natural question that arises from this work is whether our framework could be used to promote innovative economies, inhibit viral evolution, or shape scientific progress. Our starting principles lead to several relevant insights. First, some system parameters may be more opportune than others in forcing a transition in idea lattice dynamics. It may be counterintuitive that boosting growth or innovativeness or reducing obsolescence are not the most forceful ways of maintaining a diverse set of ideas but balancing death and obsolescence rates is (Eq~\ref{eq:divergence inequality}). A second intriguing prediction is that transitions from Schumpeterian dystopia and runaway innovation can be sudden (Appendix~\ref{si sec:fluctuations}). The fluctuations near the critical point highlight an opening for dynamical classification of systems through rate parameters or signaling when systems are on the verge of collapse \cite{schefferEarlywarningSignals2009}, and it may reflect endogenous dynamics that drive large-fluctuations in biodiversity \cite{mayhewBiodiversityTracks2012,rohdeCyclesFossil2005}, economic growth \cite{silverbergEvolutionaryModel1995}, or scientific decline \cite{bloomAreIdeas2020}. This puts forth the possibility of a comparative meta dynamics, where we envision tracking systems in the innovative-obsolescence space with finer-scale dynamical data. Our model organizes these hypotheses and opens up a new framework for thinking about the forces of innovation and obsolescence. After all, an engine may explode from having too much fuel or putter out from having too little, so likewise, an idea engine need be fine-tuned. 

%An analogy to markets would be to ask if incentivizing innovative spin offs or decreasing bankruptcy rates has the equivalent effect of supporting propping up the oldest firms in terms of idea diversity. Increasing replication, however, raises the threshold of obsolescence rate for collapse, potentially increasing systemic risk. %This may increase systemic risk if the parameters are coupled as they are when new investments cannibalize existing markets. 

%Another topic that we sidestepped was what kinds of things become innovations. In economics, Schumpeter distinguished between innovations, which are production mechanisms that fix in the economy, versus inventions which are fanciful ideas. In science, innovative ideas are ones that tend to connect. The reason is not necessarily that scientists are not sitting next to revolutionary ideas, but that the certain types of work may be incentivized \cite{fosterTraditionInnovation2015}.

\section*{Methods}
\subsection*{Data}
We test our model with several data sets. Here, we describe the sources of the data sets and how we calculated the values that we show in Fig.~\ref{gr:pseudogap examples}. 

The distributions of firm cost efficiency in Fig.~\ref{gr:pseudogap examples}a are digitized from Fig.~1 of reference \citenum{iwaiSchumpeterianDynamics1984} by Iwai, and the densities in Fig.~\ref{gr:pseudogap examples}b are digitized from the third panel of Fig.~3 from reference \citenum{jangiliImpactSize2019} by Jangili. In Iwai's plot, we are showing the cost of labor relative to the cost of the product such that a cost of zero implies that the value of the sale is only profit neglecting other capital costs. Iwai also notes that a similar histogram appears across industries at the time of his investigation citing Sato's 1975 publication. Iwai's interpretation of the figure aligns with the classic interpretation of firm innovativeness in that more innovative firms are the ones lowering their cost ratios \cite{aghionModelGrowth1990}. On the other hand, Jangili's distribution derives from an estimate of the relative cost efficiency of firms using a technique called stochastic frontier analysis \cite{kumbhakarStochasticFrontier2004}. In short, this technique involves estimating  the maximal cost efficiency that firms could hope to obtain from a given set of measures about firms (e.g.~size, age, liquidity, leverage, capital-labor ratio, etc.) that convey information about the costs firms incur. When cost efficiency is unity, firms have reached maximal possible efficiency. Jangili's work shows that the shape of firm distribution is relatively consistent over long periods of time, and the distribution does not drift towards perfect cost efficiency of one. Jangili's Figs.~A.1 and A.2 are of particular interest, which show nearly the same inferred distribution of firm cost efficiency over 20 years. Another example of this characteristic shape is for European financial firms \cite{badunenkoAchievingSustainable2021}. %This observations raise the question of why it is that the distributions stay roughly consistent over time. 

The classic assumption is that more efficient methods of production are more innovative because firms compete to improve their profit margins \cite{barroEconomicGrowth2003,iwaiSchumpeterianDynamics1984}. The level of aggregation, however, is important for comparing firms. As we point out in the main text, the level of aggregation is one major point of contrast between Iwai's and Jangili's analysis. As another counterpoint to the classic assumption, we note that it is at odds with the observation that disruptive firms, at first, are cost inefficient such as Tesla, which had been producing vehicles at a loss for many years \cite{alvarezTeslaVehicle2022, teslaQ4FY20212022}. In contrast, we do not start with the assumption that higher cost efficiency is more innovative, but instead have a distribution for which we seek the most reasonable mapping of the given economic variable to our model. We emphasize that our measure of innovativeness takes into account the shape of firm density, not their absolute values of cost efficiency. In this sense, we discover the appropriate axes for innovativeness in metal stamping vs.~Indian firms.

%Since firms tend to become more cost-efficient over time with an established product (the usual economy of scale argument \cite{barroEconomicGrowth2003}), we might anticipate that newer, smaller firms are the ones introducing disruptive innovations. Since newer products are often more costly to produce, this might imply that smaller firms would be less cost efficient. Instead, Jangili finds that on the whole small firms are more cost efficient than their larger competitors. The difficulty with such logic is that cost efficiency may be indirectly related to innovativeness when looking across industries as opposed to any particular one. 

%orientation of the cost efficiency axis that takes more innovative firms to be the ones that are more cost efficient. That the distribution is preserved over time seems to imply, in our framework, that the underlying lattice is drifting to the right such that older firms (and thus larger ones \cite{zhangScalingLaws2021}) become increasingly cost-efficient but also obsolete. 
%This notion of innovation would be in opposition to Iwai's, reflecting tension in the literature about how precisely to measure or to define innovation.

As is detailed in reference \cite{iwaiSchumpeterianDynamics1984}, Iwai obtains the data from Sato (1975) that were originally obtained from the U.S.~Department of Commerce. As is detailed in reference \cite{jangiliImpactSize2019}, Jangili samples 11{,}410 non-financial, public, Indian firms between the years 1995-2014 listed in the PROWESS database maintained by the Centre for Monitoring Indian Economy,\footnote{\url{https://prowessiq.cmie.com}} but the data is not publicly available. Small firms and large firms are distinguished by either belonging to the lowest or highest quartile of total assets.

We obtain the SARS-CoV-2 clade data from the Nextstrain project downloaded on August 10, 2022.\footnote{\url{https://nextstrain.org/ncov/gisaid/global}} We use the inferred phylogenetic trees based in the GISAID sequence repository that contains millions of global samples of SARS-CoV-2 strains. We focus on the European and North American subset since we expect these to be particularly well-sampled though generally it is nigh impossible to track all circulating strains. After mapping the imputed phylogenetic tree from Nextstrain into distance marked by base pair mutations, we calculate the average number of branches into which any unit length of the tree divides. In other words, a non-branching unit contributes a branch $Q-1=1$, one that leads to two clades $Q-1=2$, etc., where $Q$ is defined as the branching ratio. Taking the average number of outgoing branches, we then find the branching ratio $Q=2.317$ for North America and $Q=2.344$ for Europe. Using the calculated value, we normalize the number of unique individual clades as by $(Q-1)^y$, where $y$ is the number of mutations from the original detected strain in order to calculate the typical number of strains per branch. This will lead to an exponential decay if the number of detected strains remains constant as is approximately the case in the data far from the root strain hCoV-19/Wuhan/Hu-1/2019.

The scientific article and patent citation data in panels f and i come from the American Physical Society's repository for {\it Physical Review B} ({\it PRB}) and PatentsView (a repository for the US Patent and Trademark Office) \cite{patentsviewPatentsView2022}, respectively. For {\it PRB} citations, we consider 1{,}369 papers published in 1980 and only citations within the universe of {\it PRB} papers. We first bin these papers by total cumulative citations to the papers til 2020 as a measure of fitness. While we wish to obtain a measure of interest in a paper over time, we also know that not only is the number of annual publications increasing over time but also the length of bibliographies. To obtain a citation rate that accounts for the changes exogenous to our model, we normalize the citation counts for each paper published in 1980 by the typical number of citations made by every paper published in each following year, i.e.~we count effective citations relative to the typical number of citations made per paper per year. 

On the other hand, we separate US patents by the six technology classes that are identified in reference \cite{hallNBERPatent2002}. In Fig.~\ref{gr:pseudogap examples}, we consider the 21{,}896 patents filed in category 4 (electrical \& electronic) from 1990. We normalize the citation rates, following convention, by the number of patents filed each year within the focus technology category, a normalization factor that grows with the number of potential citation recipients within the same technology category \cite{candiaUniversalDecay2019,highamFameObsolescence2017}. This procedure again accounts for change in the population of citing patents (assumed to the proportional to the density of agents) that would not be captured in our simple model.  %We further note that this represents a more reliable way of normalizing the citation curve because there are substantial biases in the number of citations to the past in recent decades (about 20 years). These represent alternative ways of normalizing the citations to account for changes in the population that would not be captured in our model. 

%\begin{table}\centering
%\caption{Number of patents applications in given technology category in 1990 according to data acquired from PatentsView. }\label{tab:patents}
%\begin{tabular}{c|c}
%	technology category & patents \\
%\hline
%	1 (chemical)							&   23,710 \\
%	2 (computers \& comms.)					&   14,077 \\
%	3 (drugs \& medical)					& 	9,451 \\
%	4 (electrical \& electronic)			&	21,896 \\
%	5 (mechanical)							& 	27,176 \\
%	6 (others)								&   25,757 \\
%\end{tabular}
%\end{table}

%Patent citations (including those by examiners and applicants) show a similar pattern, where the number of patents increases over the years as well as the number of citations made by each patent \cite{highamFameObsolescence2017}. We again normalize citation count by the typical number of patents filed each year for the specified technology category to obtain the trajectories. As an alternative way of testing our model, we combine highly and little cited patents to obtain an estimate of the occupancy number for all patents in a given year in panel d.

The observed citation rates --- because they initially peak then slowly decay --- are consistent with the hypothesis that scientific ideas are driven by the extinction of obsolete ones. In both cases of citations we consider, the overall shapes align with our obsolescence-driven formulation. 

Our analysis implies that the majority of agents in the scientific system are closer to the innovative edge, whereas most agents are closer to the obsolescence front in the other examples. This means that science as a whole is relatively innovative. This is a result of the fact that driving an idea obsolete in the stationary case effectively makes every agent more innovative. Perversely, the same dynamic drives out the most innovative agents from the system whose exit reflects the fact that scientific experts are not particularly good at valuing novel ideas \cite{boudreauLookingLooking2016}. Firms and viral evolution, on the other hand, are forward-looking because the front is only driven ahead by most innovative agents. That means that fewer agents are near the innovative edge. %We emphasize that the exact mechanistic details implicitly depends on factors like financial viability, competitive advantage, or economic regulation. 

%Cultural evolution: A natural point of comparison with our model is cultural evolution such as is measured by word usage \cite{amatoDynamicsNorm2018}. In word usage, there are, by construction, only two possibilities, the new and the old. There is no turnover because innovation happens once by definition. If we remove that term from our equation, we get something similar to logistic dynamics proposed in Amato et al. but, . Our equation is more complicated because we have to account for the exponential decay of the old, which leads to an exponentially decaying inflow rate to the new, and a softer bend (Appendix~\ref{}).

% In order to extract patent citation curves, we considered the cohort of patents granted in the USPTO bulk data starting in the year 1990 for each technology class separately \cite{} and counted the total number of citations to the cohort from patent applications in the same and subsequent years. The granted patents are identified the ``patent'' database and we are able to extract their application information including year from the ``application'' database. We extract all citations recorded in ``uspatentcitation.'' 

\subsection*{Fitting the model}
We fit model parameters by scanning through parameter space consisting of rescaled rates for growth $G/r$, obsolescence $r_o/r$, death $r_d/r$. Innovativeness $I$ does not need to be rescaled because it is unitless. At each test parameter combination, we find the unit conversion for the lattice coordinate $a$ and the density $b$ that minimizes the linear (e.g.~firm densities) or logarithmic (e.g.~genetic and citation curves) squared distance between the data and the model as visible in Fig.~\ref{gr:pseudogap examples}. 

In order to calculate the distance, we must decide on how to align model lattice coordinates with data coordinates. It not always clear if the leftmost or the rightmost data coordinates correspond to innovation and obsolescence fronts. This problem is especially vexing in the context of firms, where one can have in principle any real-valued positive or negative cost efficiency or productivity (such as after accounting for subsidies or other external costs). In viral mutations, this question arises with establishing the ``origin'' from which the virus descends, which in principle could be traced back to the origins of life. For citations, the definition the origin is seemingly straightforward since the innovation front cannot precede the publication of the paper or patent, but again there are some practical considerations that muddy the boundary. Papers might have been cited in the first year of publication, while in preprint form, or by the authors before publication in any form. In all of these cases, we take the simplest mapping from the innovation or obsolescence front to the data. 

With firms, we take the leftmost point to correspond to the innovation front but allow the distance minimization procedure to choose the optimal location of the obsolescence front given by the scales $a$ and $b$. For SARS-CoV-2 mutations, we assume that the origin is the first detected hCoV-19/Wuhan/Hu-1/2019 strain. Since citations can occur as soon as a publication or patent filing appears, it would be natural to set the innovation front at the first year. We do so for patents. With the {\it PRB} citations, however, we find as the optimal fit with this assumption returns unusually large parameters that are difficult to handle with our grid search algorithm. Instead, by assuming that the innovation front corresponds to the year following publication such that the first year has density of zero in our model, we find similarly good fits except that the parameters are well bounded. Thus, we rely on this more controlled procedure. %While it is unclear why such regularization works, we have reason to think that the citations for papers immediately following publications may be different from citations that come later. Some PRB citations immediately following publications were presumably by others already following the work as it had been developing. This is less likely for later papers especially as time progresses. 

A secondary question is how to map the spacing of our discrete lattice to the units of the data. Since the solution corresponds to discrete values of a continuous partial differential equation, we solve the lattice solution using our flow mean-field solution, interpolate the lattice solution with a cubic spline, and use the spline values to minimize the distance between data and model.

We elucidate upon each fit in the panels of Figure~\ref{gr:pseudogap examples} beyond what has been already specified:
\begin{enumerate}
	\item [a.] Metal stamping firms: We present the solution to the histogram for 1953 upon minimizing the linear squared distance. As mentioned, the location of the obsolescence front is fit.
	\item [b.] SARS-CoV-2 clades: After using the estimated branching ratio of the phylogenetic tree $Q$ as described above, we minimize the logarithmic squared distance.
	\item [c.] Article citations: We minimize the logarithmic distance to the first decade after publication (i.e.~all citations in the years 1980 through 1990) while asserting that the best fit curve extend to at least 40 years. Thus, the later years are a predictive test of our model.
	\item [d.] Inferred cost efficiency for Indian firms: We minimize the linear squared distance. The location of the obsolescence front is fit.
	\item [g.] Patent citations: We minimize the logarithmic distance to first five years after publication (i.e.~1990-1995 inclusive) while asserting that the best fit curve must extend to at least 20 years. Thus, the later years are a predictive test of our model.
\end{enumerate}

To approximate the model occupancy densities, we rely on the second-order calculation of lattice length described in Appendix~\ref{si sec:model derivation} and calculate the density with the flow mean-field calculation in Appendix~\ref{si sec:num calc}.

\subsection*{Code availability}
The code for processing the data, calculating the model, and producing all the plots as described below is available at \url{https://github.com/eltrompetero/innovation} and will be put on an archive repository to be determined upon publication.

%\section*{Remaining items}
%\begin{enumerate}
%	\item Discuss further examples of social change with word usage
%	\item Extension to more realistic idea graphs.
%	\item Picture of innovation graphs of citations.
%\end{enumerate}

\section*{Acknowledgments}
E.D.L.~thanks Frank Neffke, Anjali Bhatt, Alan Kwan, and the CSH theory group including Tuan Pham, Jan Korbel, Stefan Thurner, Ernesto Ortega, and Rudi Hanel (who suggested the term ``innovation train'') for inspiring discussions at various stages of ideation. E.D.L.~acknowledges funding from BMBWF, HRSM 2016 (Complexity Science Hub Vienna). We thank the American Physical Society for access to journal citation data. We thank Jeff Barrick for help with the {\it E.~Coli} phylogenetic data and Am\'{e}lie Desvar-Larrive for covid-19 references.

\section*{Competing interests}
The authors declare no competing interests.

\section*{Materials \& correspondence}
Correspondence and questions about code should be sent to E.D.L.

\bibliography{refs}

\clearpage
\appendix

\setcounter{figure}{0}
\setcounter{equation}{0}
\renewcommand{\thefigure}{S\thesection.\arabic{figure}}
\renewcommand{\theequation}{S\arabic{equation}}
\renewcommand{\thetable}{S\arabic{table}}

\section{Model motivation, derivation, and solution}\label{si sec:model derivation}
\begin{sidewaystable}[b]\centering
\caption{Model dynamics and possible attributions. The provided set extends beyond the data sets considered in the main text.}\label{si tab:dynamics}
\vspace{.3cm}
\includegraphics[width=.75\linewidth]{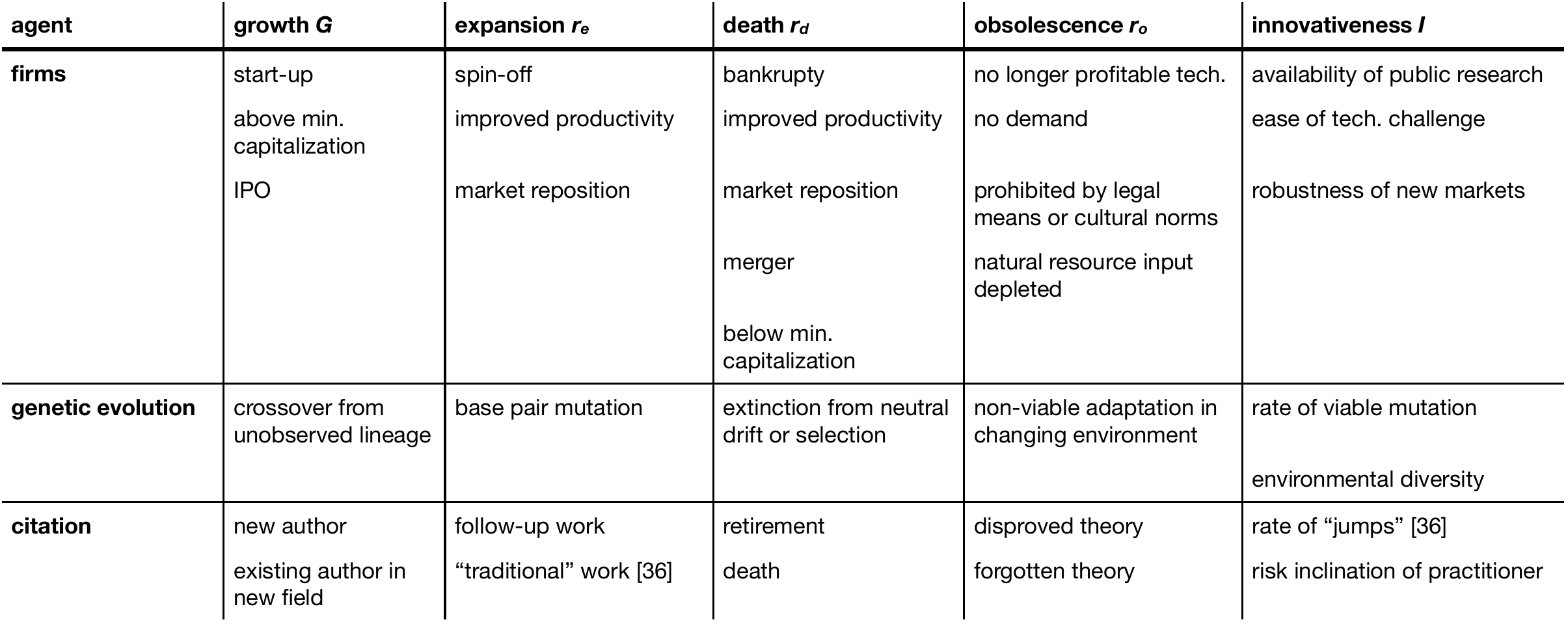}
%\resizebox{\textwidth}{!}{%
%\begin{tabular}{c|l|l|l|l|l}
%	context 	& $G$					& $r$				& $r_d$		& $r_o$	& $I$ \\
%\hline
%	firms		& start-up				& spin-off				& bankruptcy			& unprofitable technology	& difficulty of tech\\
%				& reach min. capital	& improved productivity	& improved productivity	& no market	& \ \ challenges\\
%				& public incorporation	&						& merger				&	& public research\\
%\hline
%	genetic 	& crossover from 	& base pair mutation	& selected out	& uncompetitive	& probability of viable\\
%	evolution 	& \ \ unobserved lineage &					& neutral drift	& non-viable in	& \ \ mutation \\
%				& crossover from	& 	& 	& \ \ changing environment\\
%				& \ \ other species	& 	&	&	& \\
%\hline
%	science		& new author		& follow-up work	& retirement		& disproved theory 	& jump \cite{fosterTraditionInnovation2015}\\
%				& old author in		& traditional work \cite{fosterTraditionInnovation2015}	& death				& forgotten theory\\
%				& \ \ new field		&					&					& \\
%\end{tabular}}
\end{sidewaystable}

\begin{sidewaystable}[b]\centering
\caption{Model structure and possible attributions. The provided set extends beyond the data sets considered in the main text.}\label{si tab:structure}
\vspace{.3cm}
\includegraphics[width=.75\linewidth]{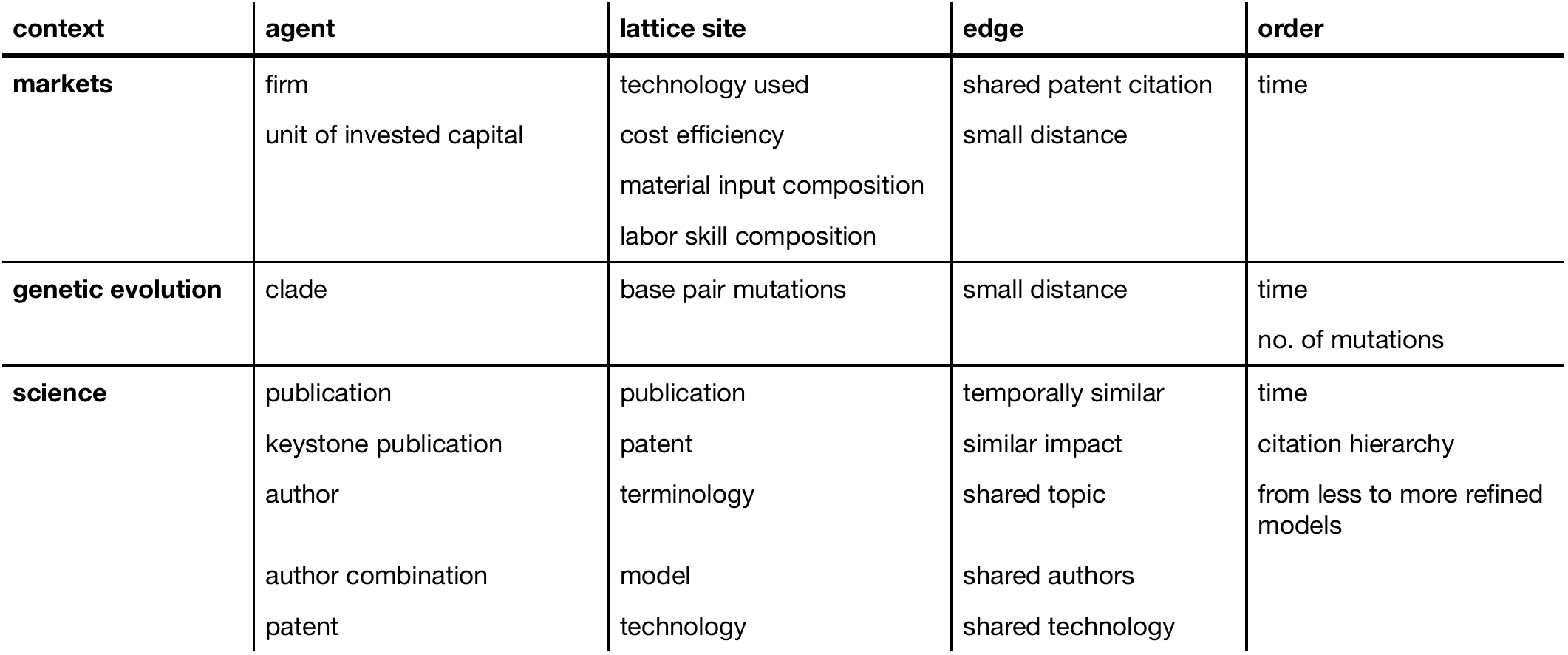}
\end{sidewaystable}

In Eq~\ref{eq:1d lattice}, we propose a general model for innovation and obsolescence dynamics of an idea lattice with replicating agents. The parameters in the model correspond to the rates at which various events occur, which could encapsulate multiple different processes at the level of our analysis. The proper mapping from each system of interest to model parameters must be operationalized, which essentially accounts for the proper units of the variables.
\begin{enumerate}
	\item The entry rate $G$ at which new agents enter the system requires defining a threshold at which we observe new agents entering. In the context of firms, we might think this to be straightforwardly the entry of a new incorporated firm in an industry. As is discussed in more detail in the study of firm demographics, however, the ``birth'' of a firm is a matter of measurement and definition. It depends on the universe of study such as listed public firms vs.~hard-to-measure private firms \cite{axtellDynamicsFirms2019}. The study of the former at the exclusion of the latter effectively establishes a minimum capital size; in other words, the effective rate at which agents enter can also depend on the level of data resolution. For viral evolution, entry captures the emergence of new strains that do not derive from existing ones, which must be from the fact that we can only observe a small fraction of circulating strains at any given time and must necessarily impute the missing pieces of the phylogenetic tree. For scientific citations, entry rate combines the rates at which new authors enter as well as that of existing authors moving into a new area.
	
	\item The rate of agent replication $r$ refers to the typical timescale at which an agents spawns a more innovative copy of itself, and the death rate $r_d$ refers to the typical timescale on which agents leave the system. Though these can be treated separately, they in combination describe various situations. As one example, $r=r_d$ corresponds to the scenario where agents hop from one site to the adjacent one because death is commensurate with a single replication event. When $r>r_d$, agents tend to stimulate more innovative copies, whereas when $r<r_d$, agents tend to die before they can replicate, akin to the scenario where agents only progress a few steps towards the innovative frontier before dying. The parameters individually can encapsulate several different types of events. For example, $r$ may be the sum of two different process like mimetic innovation and replication, but events that we would sum together into a single rate because they effectively increase the number of agents on the adjacent innovative site. Likewise, death could consist of the disappearance of an agent from a site because of mimetic innovation and death. The key assumption is that the combination of these types of events can be separately considered as mean rates that can be then summed together.
	
	\item The rate of obsolescence $r_o$ details when ideas fall into desuetude. While we explicitly write down a rate that acts on the idea lattice, usually it is considered in terms of an endogenous variable like when a production method is no longer used by firms. As we point out in the main text, not all obsolescent events can be traced back to an endogenous cause, and so this parameter is a way of representing the results of either kind of dynamics.
	
	\item Innovativeness $I$ sets a scale on which attempts at innovating into the adjacent possible are successful, distinguishing mimetic innovation from serendipitous innovation at the frontier. As with the other rates, this parameterization assumes that there is a typical rate at which innovative events occur as when new technologies are created, unseen mutations emerge, or ideas are introduced into the literature.
\end{enumerate}
When an Einsteinian innovation occurs, the parameters themselves must be changed, which violates the assumption of fixed mean rates.

Accounting for these rates means that we only need describe the number of agents $n(x,t)$ at some lattice site $x$ at time $t$. As a start, we take the innovation front to be the rightmost point in the linear lattice as pictured in Fig.~\ref{gr:model}B. First, we account for the rates entry $G$ over $L(t)$ lattice sites and the rate at which any single agent dies $r_d$. Next, the probability that at least one agent successfully advances the innovation front is $1-(1-r I dt)^{n(L,t)} = r I n(L,t)dt + \mathcal{O}(dt^2)$, where the latter term refers to all terms of order $dt^2$ and smaller. On the other hand, with probability $(1-r I dt)^{n(L,t)} \approx 1- rIn(L,t)dt$, the front does not move and $r n(x-1,t)dt$ agents move in from the left. 
%This leads to the rate of change in agent number
%\begin{align}
%\begin{aligned}
%	%\dot{n}(x,t) &= \frac{G}{L(t)} - r_d n(x,t) - rIn(L,t)n(x,t) + [1 - rIn(L,t)dt]r n(x-1,t) \\
%	\lim_{dt\rightarrow0}\dot{n}(x,t) &= \frac{G}{L(t)} - r_d n(x,t) - rIn(L,t)n(x,t) + r n(x-1,t).\label{eq:innov front}
%\end{aligned}
%\end{align}
%Eq~\ref{eq:innov front} describes a moving idea that lattice that inches forward on the right by extracting a new site from the adjacent possible and disappears on the left as sites become obsolete.
To arrive at Eq~\ref{eq:1d lattice}, we move into a fixed reference frame that does not change as the lattice moves. We reverse the coordinate system such that the innovation front is fixed at $x=0$. Flow comes in from the right and vice versa, and innovation events shift the occupancy number to the right by one. This mathematical simplifications focuses our attention on the innovation front and the shape of the pseudogap around it.

At stationarity, we obtain a formal solution to the time-independent occupancy number
\begin{align}
	n(x) &= I n(0)[n(x-1)-n(x-2)] + \frac{r_d}{r}n(x-1) - \frac{G}{r L}.\label{si eq:iter sol1}\\
\intertext{By rescaling parameters in units of $r$ as is indicated by bars, we obtain a form that reveals that replication rate sets a shared relative timescale for all the remaining parameters,}
	n(x) &= \bar r_o[n(x-1)-n(x-2)] + \bar r_d n(x-1) - \frac{\bar G}{L}.\label{si eq:iter sol2}
\end{align}	
It is the relative differences between entry, obsolescence, and death with respect to replication that distinguish the regimes of the model.

To solve for the innovation front density, we use the fact that $n(-1)=0$. We also take as an assumption $n(1) - n(0) = n(0)$, which is to say that the shape of the occupancy number about the innovation front is linear, or that the pseudogap is rather wide.
\begin{align}
%	\dot n(0,t) &= \frac{G}{L} - r_d n(0,t) - r I n(0,t)^2 + r n(1,t)\\
	\dot n(0,t) &= \frac{\bar G}{L} + (2 - \bar r_d) n(0,t) - I n(0,t)^2
\end{align}
At stationarity, we find the innovation front number to be
\begin{align}
	n(0) &= \frac{1}{2I} \left[ 2-\bar r_d + \sqrt{\left(2-\bar r_d\right)^2 + \frac{4\bar GI}{L}} \right], \label{si eq:stat n0}
\end{align}
having put aside the unphysical negative solution to the quadratic equation. Eq~\ref{si eq:stat n0} reveals that the density at the innovation front is inversely proportional to the difficulty of advancing the front to the adjacent possible $I$. Increasing the replication rate $r$ or $G$ also leads to higher innovation front density as agents pile up faster. Thus, we find simple dependence of the innovation front density on the local dynamics of agent innovation, death, and replication.

%\begin{table}\centering
%\caption{Reference table for model parameter definitions.}\label{si tab:notation}\vspace{.2cm}
%\begin{tabular}{r|l}
%variable & definition \\
%\hline
%$n$		& occupancy number or density\\
%$\dot{n}$	& time derivative of occupancy number\\
%$r_d$	& death rate\\
%$r$	& replication rate\\
%$r_o$	& obsolescence rate\\
%$t$		& time\\
%$x$		& lattice index\\
%$C$		& correction term to lattice length\\
%$G$		& entry rate\\
%$I$		& innovativeness (rescales rate of a successful attempt)\\
%$L$		& lattice length\\
%$Q$		& lattice branching ratio, $Q=2$ for linear lattice\\
%\end{tabular}
%\end{table}

%\begin{figure}\centering
%	\includegraphics[width=.6\linewidth]{n0_L_example}
%	\caption{Comparison of approximate analytical solution with automaton simulation. (top) Numerical calculation of dynamical equations with corrected lattice width from Eq~\ref{si eq:corrected L} (green) leads to better alignment with the stationary condition $n(0) = r_o (Ir)^{-1}$ (blue) compared to the na\"ive, linear calculation. Our heuristic method (red) of finding an $L$ with the iterative solution hew almost exactly the exact solution. (bottom) Comparison of (left) lattice width and (right) innovation front density for automaton and mean field theory (MFT). (right) At small $G$, the system collapses and disagrees with MFT. To the right, the start up rate $G$ is sufficiently large such that the system reaches a self-sustaining state. When obsolescence rate $r_o$ is too small (square markers), the system grows indefinitely and density exceeds MFT.}\label{si gr:mft test}
%\end{figure}

\subsection{Improving the linearized estimate for lattice width}
% commented equations have general form including Q!=2
While the linearity assumption $n(1)\approx2n(0)$ was revealing, it was meant as starting approximation. We refine the calculation of $L$, essentially solving the first-order differential equation, by considering corrections to the linear approximation. We start with the first-order ordinary differential equation (ODE)
\begin{align}
%	0 &= \frac{G}{L} + \left(\frac{r}{Q-1}-r_d\right) n(x) + r\left(\frac{1}{Q-1} - I 
	0 &= \frac{G}{L} + \left(r-r_d\right) n(x) + r\left(1 - I n(0)\right)n'(x),\label{si eq:linear form}
\end{align}
but we add on top of our local Taylor expansion to first order at $x=0$ a correction
\begin{align}
	n(1) &= 2[n(0) + \epsilon].
\intertext{This correction really comes from the fact that derivatives of higher order matter in the full Taylor series expansion, or}
	n(-1) &= n(0) - n'(0) + n''(0)/2! - \cdots.\\
\intertext{In this context, the linearization in Eq~\ref{si eq:linear form} determines the form of these higher derivatives in a compact way. Since $n(-1)=0$,}
	n'(0) &= n(0) + n''(0)/2! -n^{(3)}(0)/3! + \cdots \label{si eq:n(0)'}
%\intertext{such that}
%	n(1) &= n(0) + n'(0) + n''(0)/2 + \cdots\\
%		&= 2[n(0) + n''(0)/2 + n^{(4)}(0)/4! \cdots]\\
%	\epsilon &= 2 \sum_{k=1}^\infty \frac{n^{(2k)}(0)}{(2k)!}
\end{align}
This is a restatement of our assertion that if $n''(0)$ and all higher order derivatives are sufficiently small, then we could use a linear approximation to specify $n(x=1)$. 

Returning to Eq~\ref{si eq:linear form}, we solve for the higher order derivatives. We can do this by taking the $k$th derivative with $k\geq1$,
\begin{align}
%	0 &= \left(\frac{r}{Q-1}-r_d\right)n^{(k)}(x) + r\left[\frac{1}{Q-1}-I n(0)\right]n^{(k+1)}(x)\label{eq:deriv relation}
	0 &= \left(r-r_d\right)n^{(k)}(x) + r\left[1-I n(0)\right]n^{(k+1)}(x)
\end{align}
and the recursion relation
\begin{align}
%	n^{(k+1)}(0) &= \frac{(r-r_d)}{r[In(0)-1]}n^{(k)}(0)\\
	n^{(k)}(0) &= z^{k-1}n'(0)
\intertext{having defined}
%	z &\equiv \frac{r/(Q-1)-r_d}{r[In(0)-1/(Q-1)]}\notag\\
%		&= -\frac{1-\bar{r}_d(Q-1)}{1-\bar{r}_o(Q-1)}.
	z &\equiv \frac{r-r_d}{r[In(0)-1]}\notag\\
		&= -\frac{1-\bar{r}_d}{1-\bar{r}_o}.
\end{align}
This means that our approximation for linearity at the front is also the condition that $z\ll 1$, an approximation that fails as $1 - \bar r_o\rightarrow 0$ for finite $1-\bar r_d$. We show the effects of this factor in terms of the expected stationary value at $n(0)$ in Fig.~\ref{si gr:n0 corrections}.

\begin{figure}\centering
	\includegraphics[width=\linewidth]{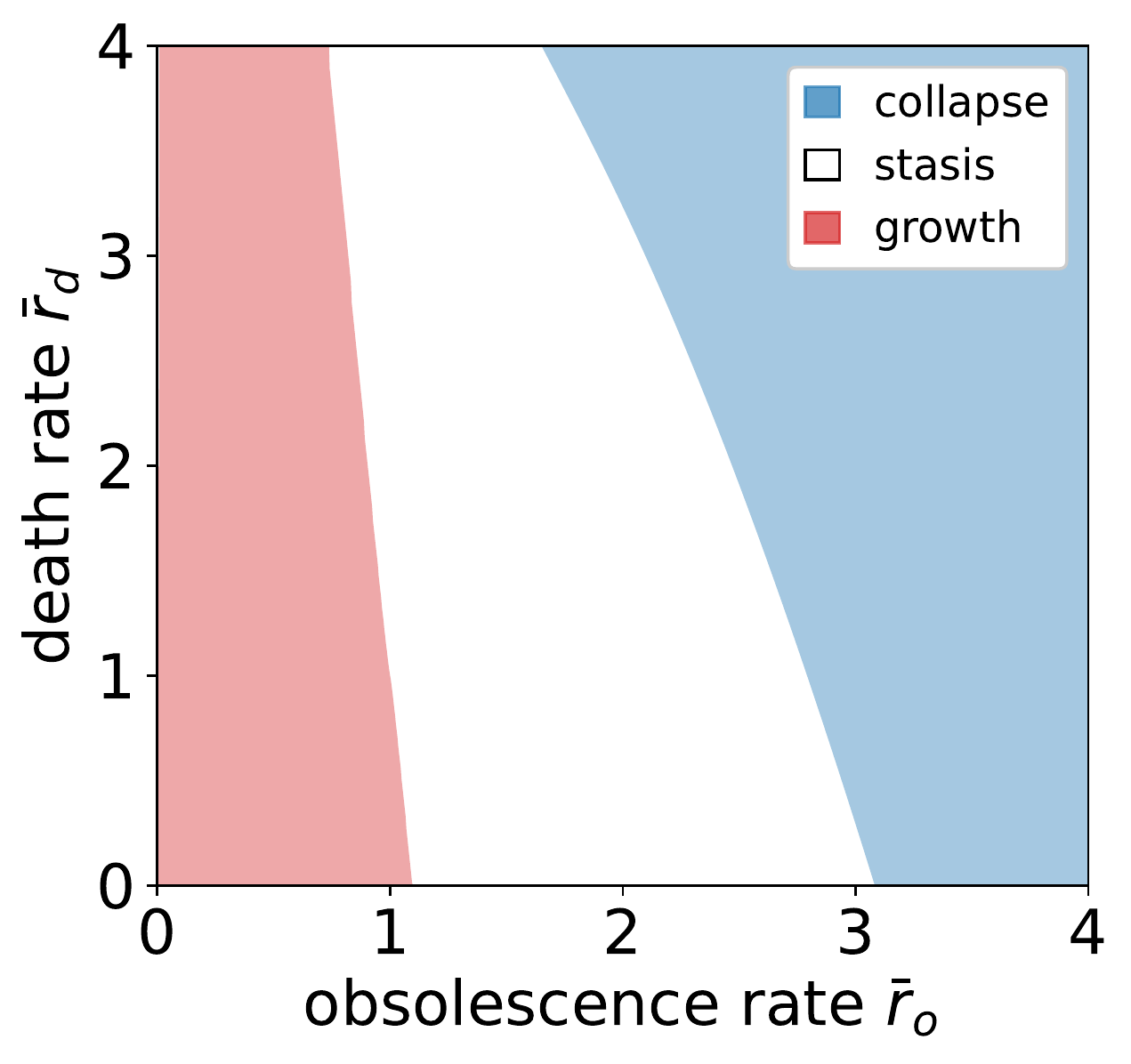}
	\caption{Phase space according to solution to the first-order ODE in Eqs~\ref{si eq:corrected L} and \ref{si eq:1st order L 2}.}\label{si gr:corrected L phase space}
\end{figure}

Putting these calculations together, the corrected pseudogap slope from Eq~\ref{si eq:n(0)'} is
\begin{align}
\begin{aligned}
	n'(0) &= n(0) + n'(0) z^{-1}\left[ \frac{z^2}{2} - \frac{z^3}{3!} + \cdots \right]\\
		&= n(0) + n'(0) z^{-1}\left[ e^{-z} - 1 + z \right]\\
	n'(0) &= \frac{n(0)}{1 - C}\\
	C &\equiv (e^{-z}-1+z)/z.
\end{aligned}
\end{align}
Going back to the stationary condition in Eq~\ref{si eq:linear form}, we can solve for $L$ but now accounting for this correction
\begin{align}
%	L &= -\bar G\left(n(0)\left[\frac{1}{Q-1} - \bar r_d + \frac{1}{1-C}\left(\frac{1}{Q-1}-In(0)\right)\right]\right)^{-1}.\\
	L &= -\bar G\left(n(0)\left[1 - \bar r_d + \frac{1}{1-C}\left(1-In(0)\right)\right]\right)^{-1}.\\
\intertext{Replacing $n(0)$ with the stationary condition for innovation front occupancy}
%		&= \bar GI\left(\bar r_o\left[\bar r_d+\frac{\bar{r}_o}{1-C} - \frac{1}{Q-1}\left(1+\frac{1}{1-C}\right)\right]\right)^{-1}.\label{si eq:corrected L}
		&= \bar GI\left(\bar r_o\left[\bar r_d+\frac{\bar{r}_o}{1-C} - \left(1+\frac{1}{1-C}\right)\right]\right)^{-1}.\label{si eq:corrected L}
\end{align}
In order to check what happens to the singularity where $L$ diverges with $\bar r_o$, we check when the denominator of Eq~\ref{si eq:corrected L} goes to 0,
\begin{align}
%	\bar r_o &= (1-C)\left[\frac{1}{Q-1}\left( 1+\frac{1}{1-C} \right) - \bar r_d\right].
	\frac{\bar r_o}{1-C} &= \left( 1+\frac{1}{1-C} \right) - \bar r_d,
\end{align}
where $C$ depends implicitly on both $\bar r_d$ and $\bar r_o$. Eq~\ref{si eq:corrected L} modifies the boundaries of the phase space as graphed in Fig.~\ref{si gr:corrected L phase space}.

\subsection{First-order solution}\label{si sec:1st ode}
We can also take a direct approach to solving the first-order ODE by simply integrating Eq~\ref{si eq:linear form}. Then, 
\begin{align}
%	0 &= \frac{\bar G}{L(t)} + (1-\bar r_d)n(x) + (1-\bar r_o)n'(x)\\
%	n'(x) &= -\frac{1-\bar r_d}{1-\bar r_o}\left[ n(x) + \frac{\bar G}{L(1-\bar r_d)} \right]\\
	n(x) &= \left(\frac{\bar r_o}{I} + \frac{\bar G}{L(1-\bar r_d)} \right) \exp\left( -\frac{1-\bar r_d}{1-\bar r_o}x\right) - \frac{\bar G}{L(1-\bar r_d)},
\end{align}
having used the starting assumption $n(0) = \bar r_o/I$. This approximation shows clearly that the slope of the exponential rise at $x=0$ and its concavity are determined by the competition between death $\bar r_d$ and competition $\bar r_o$.

We obtain an a simple form for lattice length by using the boundary condition $n(x=-1)=0$,
\begin{align}
	L &= \frac{\bar G}{1-\bar r_d}\frac{I}{\bar r_o} \left[ \exp\left( -\frac{1-\bar r_d}{1-\bar r_o} \right)-1\right].\label{si eq:1st order L 2}
\end{align}
Eq~\ref{si eq:1st order L 2} is equal to Eq~\ref{si eq:corrected L}, but the latter gives a form that aligns with the derivation presented in the main text.

We note that lattice length is not a physical quantity in the first-order ODE because the occupancy function never intersects with the $x$-axis beyond $x=-1$. This observation suggests that we must at least go to a second derivative of $n(x)$ before we can expect to approximate well the occupancy function.

\subsection{Second-order solution}\label{si sec:2nd ode}
A solvable, continuum formulation that captures the rise and drop required for a finite lattice would be helpful for solving for lattice length $L$. Here, we present a second-order ordinary differential equation approximation for an interpolation through the iterative solution in Eq~\ref{si eq:iter sol1}. Expanding the discrete formulation about $x$, we obtain
\begin{align}
	\dot n(x, t) &= \frac{G}{L(t)} + (r-r_d)n(x) + r[1-In(0)]n'(x) + \notag\\
		&\qquad\qquad \frac{r}{2}[1+In(0)]n''(x),\label{si eq:2nd ode1}
\end{align}
where we have stopped at second-order to leverage the two conditions, $n(0)=r_o(r I)^{-1}$ and $n(-1)=0$. Replace $n(0)$ with the stationary condition and taking rescaled rates, we obtain
\begin{align}
	0 &= \frac{\bar G}{L(t)} + (1-\bar r_d)n(x) + [1-\bar r_o]n'(x) + \frac{1}{2}[1+\bar r_o]n''(x).\label{si eq:2nd ode2}
\end{align}
Eq~\ref{si eq:2nd ode2} is an inhomogeneous second-order differential equation that we can solve using the standard method of variation of coefficients after having solved the homogenous equation. Since the inhomogenous term is a constant, the particular solution is likewise a constant. Then, the particular solution with characteristic eigenvalues $\lp$ and $\lm$ from the homogenous equation is
\begin{align}
\begin{aligned}
	n(x) &= A e^{\lp x}  + B e^{\lm x} + C\\
	\lambda_\pm &= \frac{1}{1+\bar r_o}\left( \bar r_o - 1 \pm \sqrt{(1-\bar r_o)^2 -2(1-\bar r_d)(1+\bar r_o}) \right).
\end{aligned}\label{si eq:ode sol1}
\end{align}
By using Eqs~\ref{si eq:2nd ode2} and \ref{si eq:ode sol1}, we find
\begin{align}
	C &= -\frac{\bar G}{L(1-\bar r_d)}.
\end{align}
We then solve for $A$ and $B$ using the boundary conditions. 
\begin{align}
\begin{aligned}
	A &= \frac{\frac{G e^{\frac{\bar r_o-1-\sqrt{z}}{1+\bar r_o}}}{L (1-\bar r_d)}-\left(\frac{2 G (1+\bar r_o)}{L \left[(1-\bar r_o)^2-z \right]}+\frac{\bar r_o}{I}\right)}{e^{-\frac{2 \sqrt{z }}{1+\bar r_o}}-1}\\
	B &= \frac{\frac{G e^{\frac{\bar r_o-1+\sqrt{z}}{1+\bar r_o}}}{L (1-\bar r_d)}-\left(\frac{2 G (1+\bar r_o)}{L \left[(1-\bar r_o)^2-z \right]}+\frac{\bar r_o}{I}\right)}{e^{\frac{2 \sqrt{z }}{1+\bar r_o}}-1}\\
	z &= (1-\bar r_o)^2 - 2(1-\bar r_d)(1+\bar r_o).
\end{aligned}
\end{align}

Eq~\ref{si eq:ode sol1} provides us two potential consistency equations for solving for $L$. Since we know that {$n(-1)=0$}, we have the following transcendental equation that must be satisfied:
\begin{align}
	L &= \frac{\bar G}{1-\bar r_d}\left(A e^{-\lp} + B e^{-\lm}\right)^{-1}.\label{si eq:consistent L 1}
\end{align}
Alternatively, we have the condition $n(L)=0$,
\begin{align}
	L &= \frac{\bar G}{1-\bar r_d}\left(A e^{\lp L} + B e^{\lm L}\right)^{-1}.\label{si eq:consistent L 2}
\end{align}
Eqs~\ref{si eq:consistent L 1} and \ref{si eq:consistent L 2} do not in general return the same result for $L$, but we find that the latter agrees better with automaton simulations. For some parameter regimes, this condition is numerically degenerate for $L$. In these cases, we rely on either the heuristic solution for the iterative equation as discussed in Appendix~\ref{si sec:iterative sol} or the linear approximation discussed in the main text.

\begin{figure}\centering
	\includegraphics[width=\linewidth]{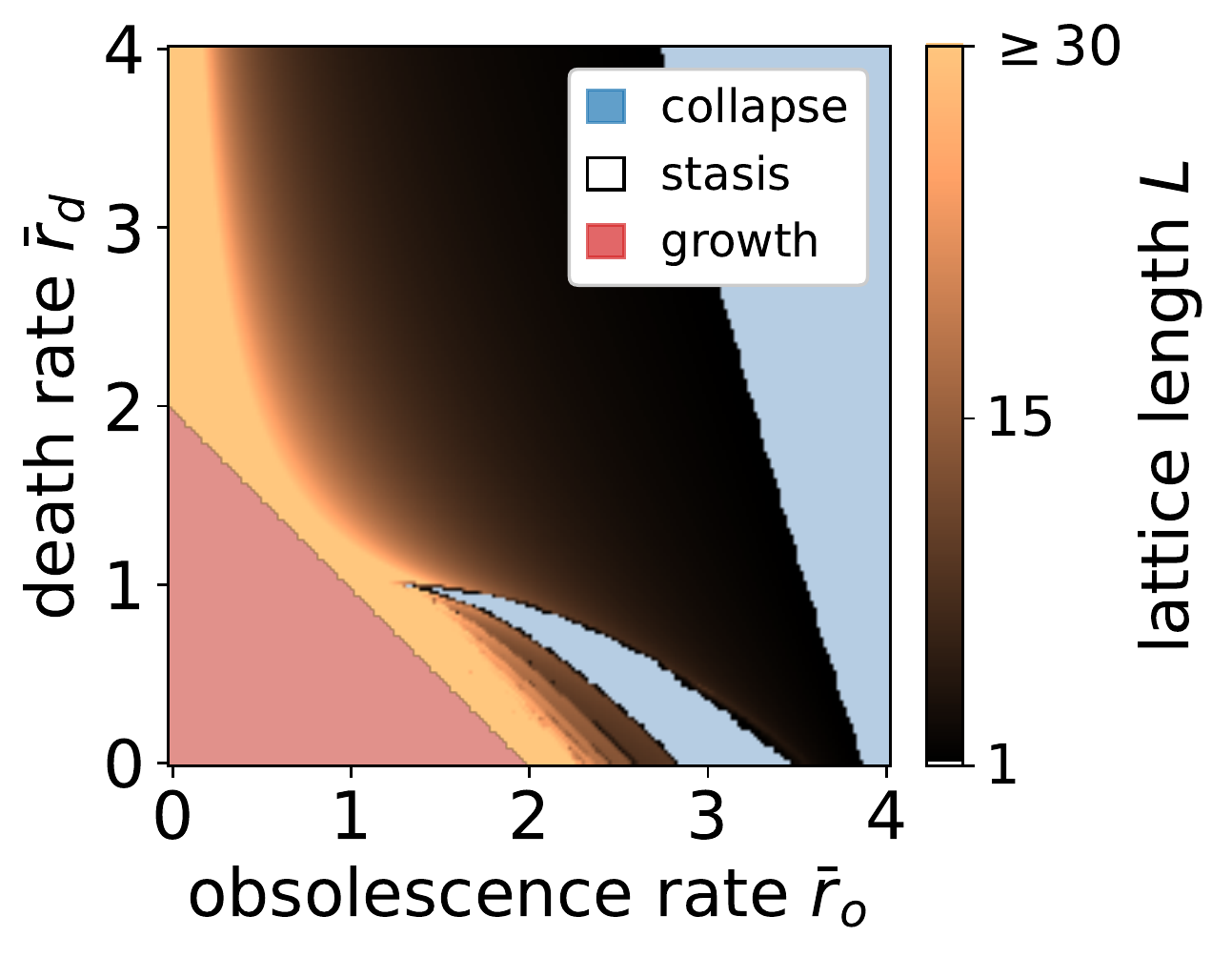}
	\caption{Numerical calculation of phase space from refined, second-order differential equation approximation in Appendix~\ref{si sec:2nd ode}. Solved by setting the initial condition to the linear approximation in Eq~\ref{eq:L} and the self-consistency condition in Eq~\ref{si eq:consistent L 2}. Static region is shaded such that darker is a longer lattice as indicated by the color map. Critical point at $r_o=r_d=r$ shows different behavior depending on angle of approach. For example, we anticipate from the first-order solution that concavity of the curve at the pseudogap depends on the ratio $(1-\bar r_d)/(1-\bar r_o)$, which can diverge, be negative, or be positive depending on the limit taken towards the critical point.}\label{si gr:2nd ode phase space}
\end{figure}

\section{Numerical calculations}\label{si sec:num calc}
We calculate the occupancy number using the iterative form of Eq~\ref{si eq:iter sol1}, the analytic approximation presented in Appendix~\ref{si sec:2nd ode}, an automaton simulation of agents, and a mean-field flow implementation of the dynamics in Eq~\ref{eq:1d lattice}. Each of these calculations have respective weaknesses and strengths that we discuss briefly in the following bullet points. More details about the implementation follow, and the code for each solution is located in the aforementioned repository.
\begin{itemize}
	\item While it is exact, the iterative calculation can be subject to numerical precision errors that grow with the lattice coordinate. It also requires a precise estimate of $L$, without which it may diverge or return negative values. Because of such divergence, the iterative calculation does not always permit an easy estimate of $L$ using self-consistency of the equation.
	\item The analytic solution, on the other hand, provides a ready way to estimate $L$ by using the consistency condition in Eq~\ref{si eq:consistent L 1}. It is an excellent approximation to the iterative solution for small densities and lattice widths, but its accuracy decreases as higher derivatives become increasingly important far from the pseudogap and for larger values of $n$.
	\item The automaton model allows us to access non-stationary dynamics and stochastic fluctuations, but it is relatively slow to calculate. It only gives an stochastic estimate of the lattice width $L$, which is necessary to condition upon to compare with the mean-field models of occupancy number.
	\item Finally, the dynamical mean-field flow calculation couples flow of agent density between adjacent lattice sites. This approach obtains temporal trajectories like with the automaton model but is much faster. It is, however, subject to discrete lattice corrections and requires knowledge of $L$ beforehand.
\end{itemize}
We leverage the set of alternative calculations to shed light on corrections and details that arise with each respective calculation. Some of the limitations may be important to consider when applying our results to practice. We describe the calculations in more detail below.

\subsection{Iterative calculation}\label{si sec:iterative sol}
Assuming that we have an estimate of lattice length $L$, we start with the stipulation $n(-1) = 0$ and $n(0) = r_o(r I)^{-1}$ to calculate the occupancy number up to $x=\lfloor L-1\rfloor$. Note that we set the innovation front to be at $x=L-1$ and $n(x=L)=0$ such that the lattice consists of $L$ sites.

A heuristic approach to estimating the lattice length for a given set of parameter relies on the observation that agent number must be finite and positive when stationarity. We run the iterative solution starting with the initial estimate for $L$ from linearized mean-field theory. When $L$ is slightly to large, the function diverges quickly to infinity and when it is slightly too small it quickly diverges to negative infinity. Using the wiggling of the tail, we can in many cases narrow our estimate of $L$ precisely. We do not expect $L$ to always depend so sensitively on errors this heuristic technique is not guaranteed to work.

\subsection{Analytic solution}
The solution to the second-order approximation of the difference equation presents a self-consistency condition for $L$. We use a standard Broyden–Fletcher–Goldfarb–Shanno (BFGS) minimization algorithm implemented in NumPy to solve for $L$ starting with the initial condition given by the linearized mean-field approximation.

\subsection{Automaton model}
We simulate the dynamics with an automaton model that consists of individual agents that follow the rules stipulated in Eq~\ref{eq:1d lattice} but with a moving coordinate system (instead of with an innovation front fixed at $x=0$). At a given step in time, each individual agent dies with probability $r_ddt$, replicates towards the innovation front with probability $rdt$, and the agents on the front each extend with probability $r I dt$. The obsolescence front progresses with probability $r_o dt$. In the limit $dt\rightarrow0$, each of these steps commutes with the others (the non-commuting corrections are of lowest order $dt^2$). Furthermore, we no longer need to consider each agent separately but just the sum of the rates of all the agents on any site $n(x)$. Thus, the small time step limit allows us simplify the calculation as is specified in the code repository and summarized below.

Assuming that $x=-1$ is the obsolescence front and $x=L-1$ is the innovation front, one algorithm is
\begin{enumerate}
	\item Instantiate lattice of size 1.
	\item Extend right side of lattice at $x=L(t)-1$ by unit length with probability $r I n(L-1) dt$.
	\item Remove a lattice site from the left hand side with probability $r_o dt$ unless lattice is already of size 1.
	\item For each lattice site $x$ in the order of the rightmost to the leftmost do the following:
		\subitem a. If $x<L(t)-1$, add a new agent to site $x+1$ with probability $r n(x)dt$.
		\subitem b. Remove one agent with probability $r_d n(x)dt$.
		\subitem c. Add one agent with probability $G\,dt/ L(t)$.
	\item Return to Step 2.
\end{enumerate}
Note that variations of this algorithm give exactly the same results in the limit $dt\rightarrow 0$ though some may converge to the limit slower than others with $dt$. The important point is that terms of order $dt$ be preserved in the calculation. 

\begin{figure}\centering
	\includegraphics[width=\linewidth]{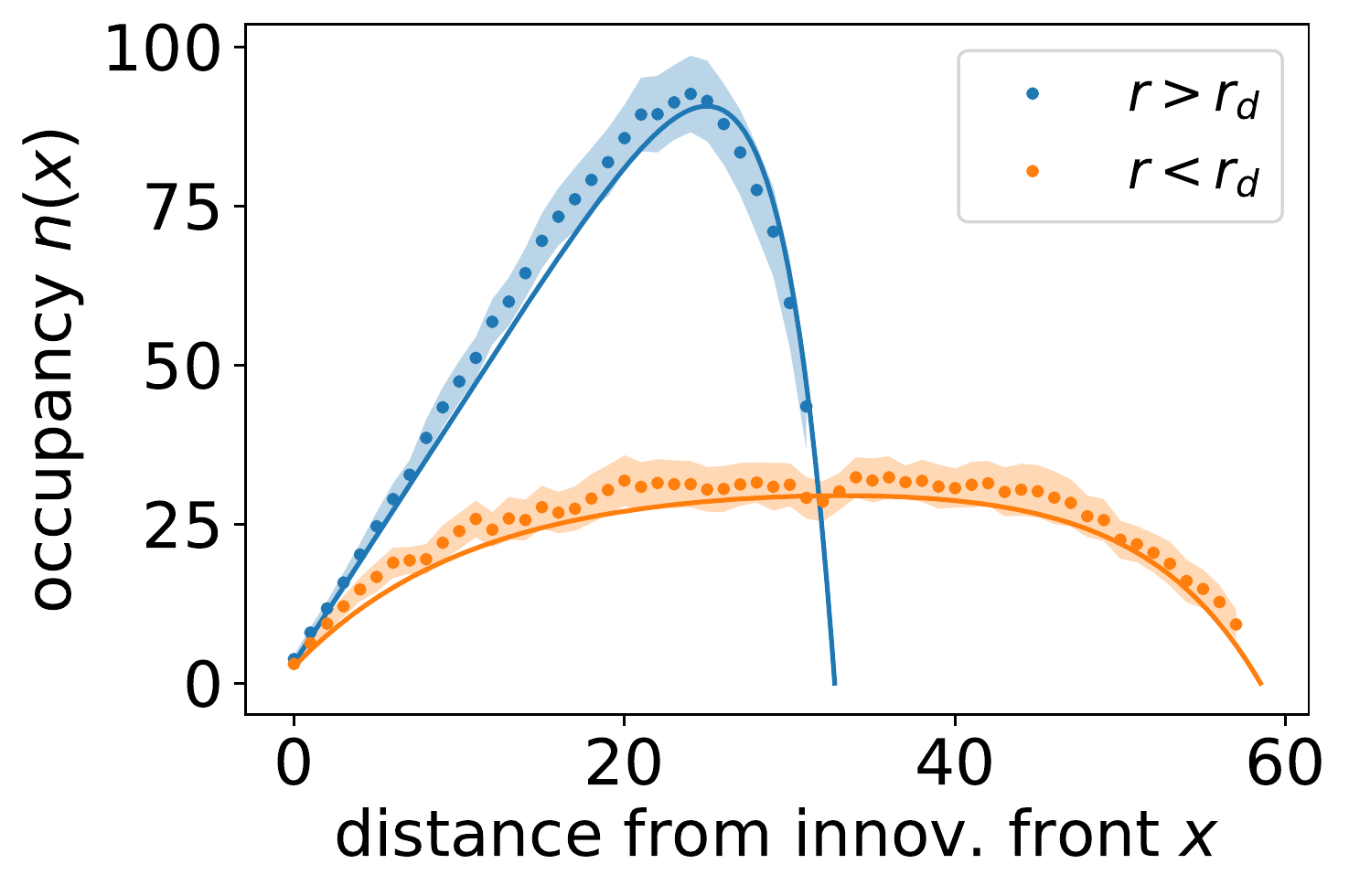}
	\caption{Comparison of automaton simulation (markers) with rescaled mean-field theory (lines). Error bars represent two standard deviations over independent lattice trajectories that have length close to the stable configuration from $L-1$ to $L+1$. Parameters are specified in the code repository.}\label{si gr:auto rescaling}
\end{figure}

The mean-field model that we discuss will not generally agree with the time or ensemble averaged occupancy number of automaton simulations. This is because the mean-field is based on the assumption that the average lattice width is the same as the average of its inverse. From the Cauchy inequality, however, we know that the typical width of the lattice in the automaton simulation will always be smaller, effectively mapping the automaton model to another mean-field equation. In order to align the mean-field theory with automaton results, we recognize that it is necessary to rescale $L\rightarrow c L$ and $x\rightarrow cx$ in Eq~\ref{si eq:iter sol1}. This transformation corresponds to the rescaling $G\rightarrow cG$ and $I\rightarrow cI$. A second factor that we must account for is the variation about $L$. To obtain an averaged occupancy number function for comparison, we restrict the average to the snapshots that are close to the stable value of $L$. In this way, we simply average over the occupancy number in the lattice relative to the innovation front up to the length of the shortest considered lattice (unless specified to the contrary). We show just one example of such a comparison that accounts for these two corrections in Fig.~\ref{si gr:auto rescaling}, which shows close alignment between automaton and the corresponding mean-field calculation.

\subsection{Mean-field flow}
\begin{figure}\centering
	\includegraphics[width=\linewidth]{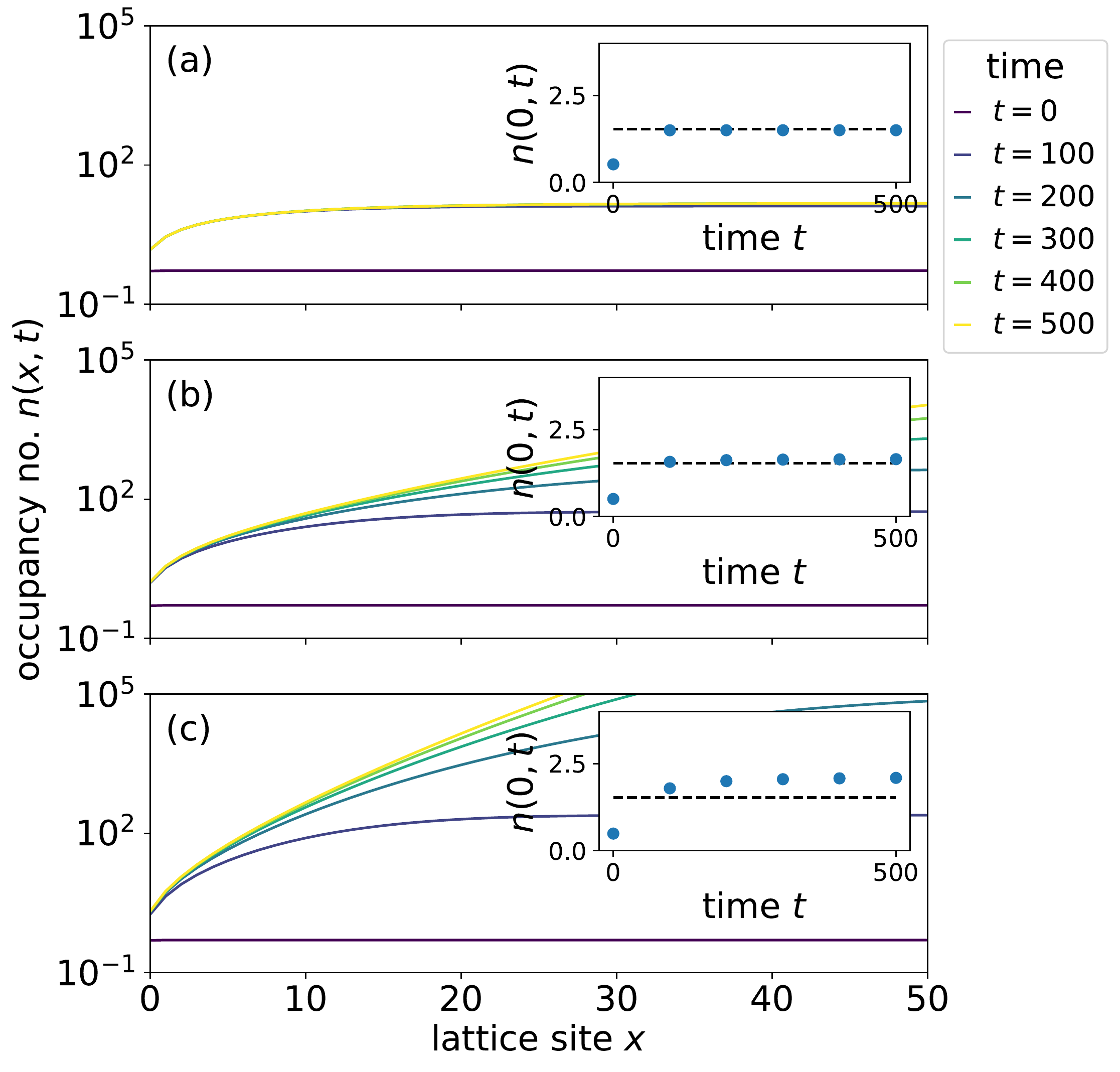}
	\caption{Simulated dynamics from mean-field flow. Numerical solution is close to expected density at the innovation front when (a) the stationarity condition is fulfilled for $r<r_d$ and (b) $r>r_d$ but not when (c) stationarity is violated in the growing regime.}\label{si gr:cont}
\end{figure}

\begin{figure}\centering
	\includegraphics[width=\linewidth]{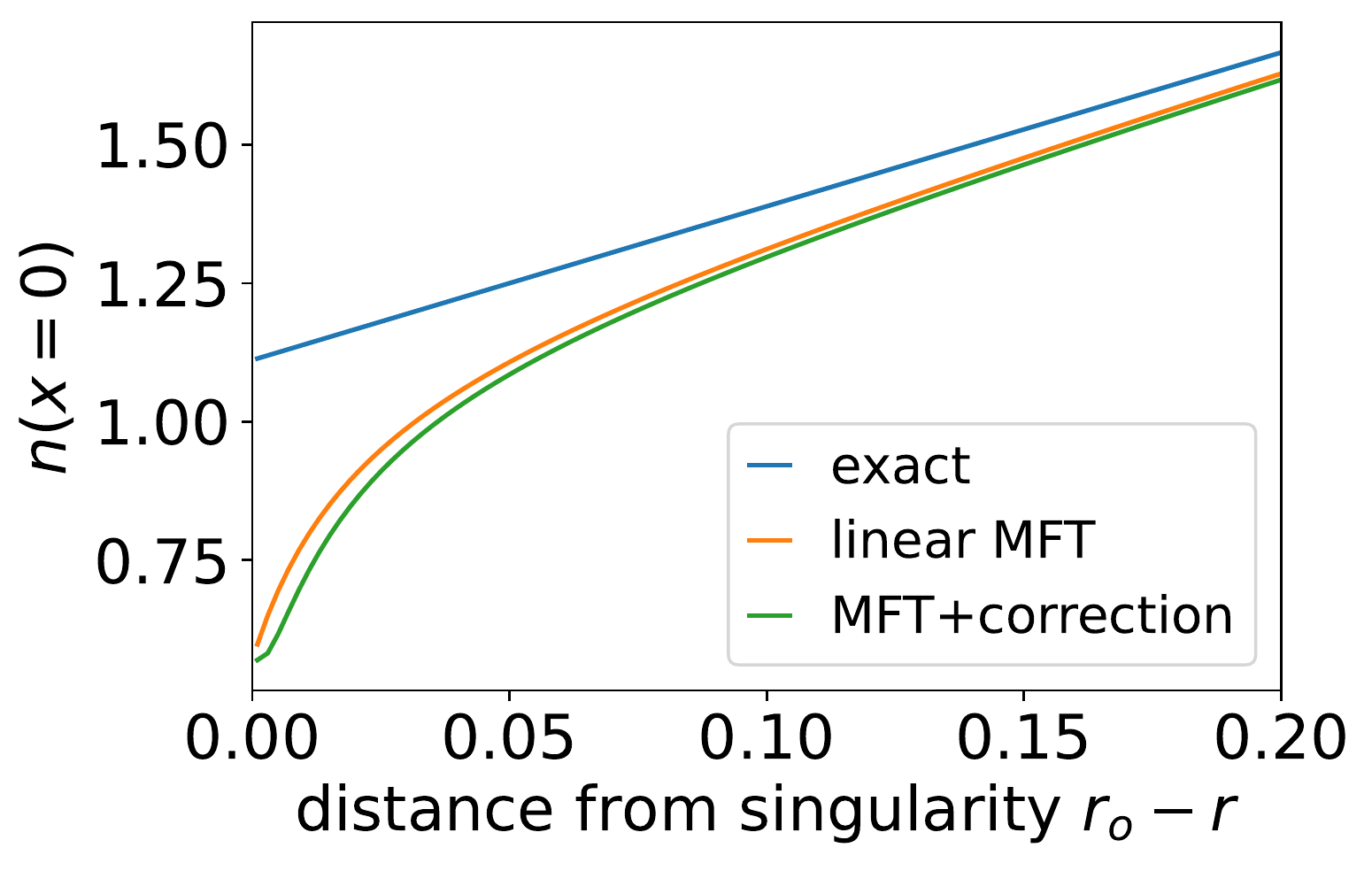}
	\caption{Comparison of innovation front number $n(0)$ approximations for different mean-field theories (MFT).}\label{si gr:n0 corrections}
\end{figure}

Having fixed the lattice size to be $L$ using the algorithm specified for the analytic calculation, we couple adjacent site densities from Eq~\ref{eq:1d lattice} and evolve the densities in time increments of $dt$. We stop the iteration when a convergence criterion has been met for the maximum absolute change $\dot n(x, t)$. This allows us to track the evolution of the occupancy number function from any initial condition as in Fig.~\ref{si gr:cont}.

\section{Model extensions}\label{si sec:extensions}
\subsection{Cooperative innovation}
\begin{figure}\centering
	\includegraphics[width=\linewidth]{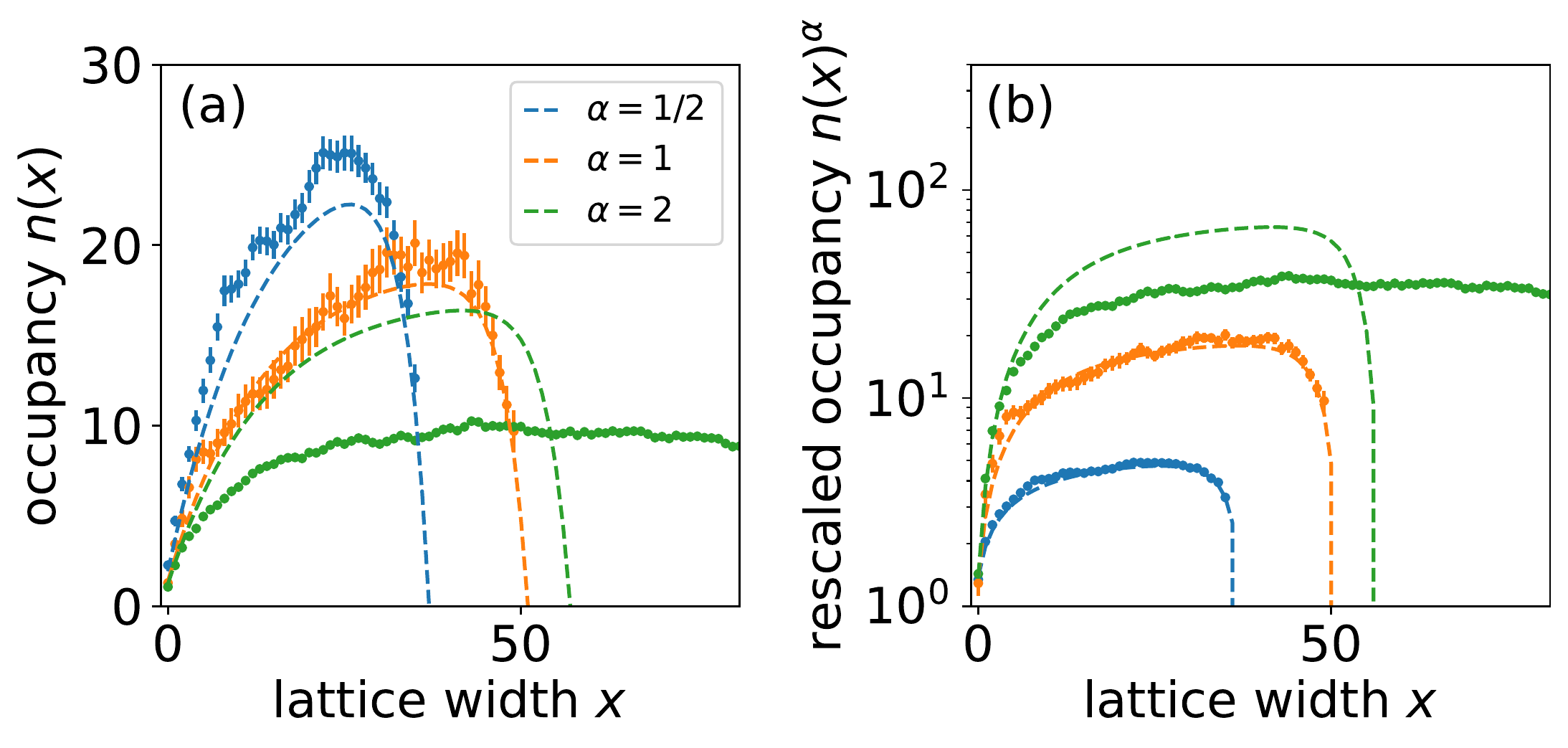}
	\caption{Examples of cooperative $\alpha=1/2$, competitive $\alpha=1$, and anti-cooperative $\alpha=2$ occupancies for mean-field theory (dashed line) and automaton simulation (marker). (b) Under appropriate rescaling, the innovation fronts collapse. For $\alpha=1/2$ and $\alpha=1$, we take an average over simulations where $L$ is close to the mean-field estimate, but mean-field theory does not accurate estimate the divergence in lattice width for $\alpha=2$. Error bars show standard deviation over independent trajectories though they are small in panel b.}\label{si gr:cooperativity}
\end{figure}

Cooperative innovation implies that front velocity scales nonlinearly with the number of agents, or that $r I n(0)^\alpha$ for $\alpha\in[0,\infty]$. The stationary condition is now $n(0)^\alpha = r_o/rI$. When $\alpha>1$, we have cooperative behavior, and $\alpha<1$ implies anti-cooperativity. The nonlinearity means that the innovation fronts coincide under the rescaling $n(x)^{1/\beta}$ as in Fig.~\ref{si gr:cooperativity}. Modifying Eq~\ref{eq:1d lattice} accordingly, we find that $L = G[n(0)(r_d + rIn(0)^\alpha - 2r)]^{-1}$, corresponding to the phase boundaries shown in Fig.~\ref{gr:phase space}b and c. The boundary separating stasis from growth is
\begin{align}
\begin{aligned}
%	\bar r_o^\alpha I^{1-\alpha} &\leq 2 - \bar r_d\\
	\bar r_o &\leq (2-\bar r_d)^{1/\alpha}I^{(\alpha-1)/\alpha}
\end{aligned}
\end{align}
whereas the boundary separating stasis from collapse requires solving for the zeros of the fractional polynomial
\begin{align}
	\bar G &= (\bar r_o/I)^{1/\alpha} (\bar r_d + \bar r_o - 2).
\end{align}
With the transformation $I^\alpha\rightarrow I$ and $(r_o/r)^\alpha\rightarrow r_o/r$, the stationary cooperative equation maps back to Eq~\ref{eq:1d lattice}. Thus, cooperativity does not alter the qualitative aspects of the model.

\subsection{Bethe lattices and higher dimensions}\label{si sec:bethe}
Bethe lattices and Euclidean graphs of higher dimension imply that the number of agents per site decreases every sequential step from the origin. In the Bethe lattice picture from Fig.~\ref{gr:model}c, each sequential site branches into $Q-1$ additional branches. A new agent must choose a branch and thus the replication term from Eq~\ref{eq:1d lattice} acquires a factor of $(Q-1)^{-1}$, and agent number decreases towards the innovation front faster than in a linear graph. As we show in Fig.~\ref{gr:phase space}b, collapsed and stable regimes grow larger at the expense of the growing regime. This argument makes clear the importance of the relative dimensions of agent replication and the idea space: when next-generation agents do not fill all of the available space, then agents necessarily occupy a small fraction of the idea space.

To make this explicit, we show the transformed equations for the Bethe lattice. Assuming that every step further out into the adjacent possible opens up $Q$ possibilities, where $Q=2$ corresponds to the linear case, then the density per site should decrease per branching point as agents decide on a branch to take. The site-specific dynamics --- including death, startup, and shift --- remain the same. To treat each branch as a replica of the one-dimensional case, we assume that the entry rate term takes the same form and decays inversely only with lattice length, which means that the total number of incoming agents scales with the number of branches. Then, the analogous equation to Eq~\ref{eq:1d lattice} is
\begin{align}
\begin{aligned}
	\dot n(x,\bar t\,) &= \frac{\bar G}{L(t)} -\bar r_d n(x,t) + \frac{1}{Q-1} n(x+1,t) - \\
		&\qquad\qquad In(0,t)[n(x,t)-n(x-1,t)].\label{si eq:bethe lattice}
\end{aligned}
\end{align}
We have rescaled time in units of the replication rate $\bar t\equiv rt$. Eq~\ref{si eq:bethe lattice} is the dynamical equation for a Bethe lattice, where $L(\bar t\,)$ refers to the number of branching steps (not sites) between the innovation and obsolescent fronts at time $\bar t$. 

The stationary solution is
\begin{align}
\begin{aligned}
	n(x) &= (Q-1)\bigg\{I n(0)[n(x-1)-n(x-2)] + \\
	&\qquad\qquad\qquad \left.\bar r_d n(x-1) - \frac{\bar G}{L}\right\}.
\end{aligned}
\end{align}
Then, the divergent growth condition is $0\leq \bar r_o \leq 2/(Q-1) - \bar r_d$, and the collapse condition from Eq~\ref{eq:obs rate} is $\bar r_o \sim (Q-1)^{-1} -\bar r_d/2 + \sqrt{((Q-1)^{-1}-\bar r_d/2)^2 + \bar GI/L}$. The dimensional depletion effect does not fundamentally alter the dynamics. If we rescale $r\rightarrow (Q-1)r$ and $I\rightarrow I/(Q-1)$, then we recover Eq~\ref{eq:1d lattice}. 

In a similar sense, we consider how the local density of agents living in a $d$-dimensional space must divide themselves as they move outwards from the origin at distance $x$. The local density decreases as
\begin{align}
	\frac{(x-1)^{d-1}}{x^{d-1}} &= (1-1/x)^{d-1}\\
	&\approx 1-(d-1)1/x
\end{align}
Clearly, the local curvature of the surface determines how thinly agents have to spread themselves out as they move further out. In the limit of a long-running lattice, $x\rightarrow\infty$, and finite dimension, this case reduces to the one-dimensional model. Thus, considering a tree-like lattice or higher dimensions does not appreciably alter the basic model.

\subsection{Obsolescence-driven innovation}
Obsolescence-driven innovation is the antithesis of forward-looking innovative systems. By reversing the direction of the $x$-axis and fixing the new innovation front at the origin, agents now replicate towards obsolescence. For stable pseudogaps, the introduction of a new idea drives every agent to left, making the whole system more innovative. As a result, innovation is driven at a rate proportional to the number of ideas on the verge of being extinguished. Newer ideas tend to beget agents on older ideas with rate $r$, which then drive themselves to extinction by eventually increasing occupancy at the obsolescence front.

We have a mirrored version of Eq~\ref{eq:1d lattice}
\begin{align}
\begin{aligned}
	\dot n(x,\bar t\,) &= \frac{\bar G}{L(\bar t\,)} - \bar r_d n(x,\bar t\,) - n(x-1,\bar t\,) + \\
		&\qquad\qquad\qquad I n(0,t)[n(x,\bar t\,)-n(x+1,\bar t\,)].
\end{aligned}
\end{align}

Let us consider the stationary case. Since sites are now added at the obsolescent front, the addition of a new site shifts all the agents one lattice site towards the innovation front. In other words, an innovation in this system means that all agents move to the left, introducing a new site adjacent to the obsolescence front and removing all agents previously at the innovation front. The removal of agents at the innovation front effectively imposes an threshold below which we are unable to detect innovative agents at stationarity. These agents beyond the known innovative frontier do not contribute any longer to lattice dynamics.\footnote{Even in the original formulation, an assumption of a continuous density function that smoothly goes to 0 would imply that some agents exist in the innovative adjacent possible and in the obsolescent adjacent possible.} We might interpret this as the fact that innovative ideas are around in some form before they are measured, so this establishes a threshold above which we recognize an innovative idea.

This threshold is linked to the obsolescence rate $r_o$, which determines how quickly the most innovative agents leave the system. This process, the disappearance of the most innovative agents,  either is slow enough for lattice growth or sufficiently rapid for collapse. In this sense, it is more appropriate to call the obsolescence rate an indulgence of unconventional ideas, which, in the stationary case, is equal to the innovation rate.

\section{Critical lines}\label{si sec:fluctuations}
\begin{figure}\centering
	\includegraphics[width=\linewidth]{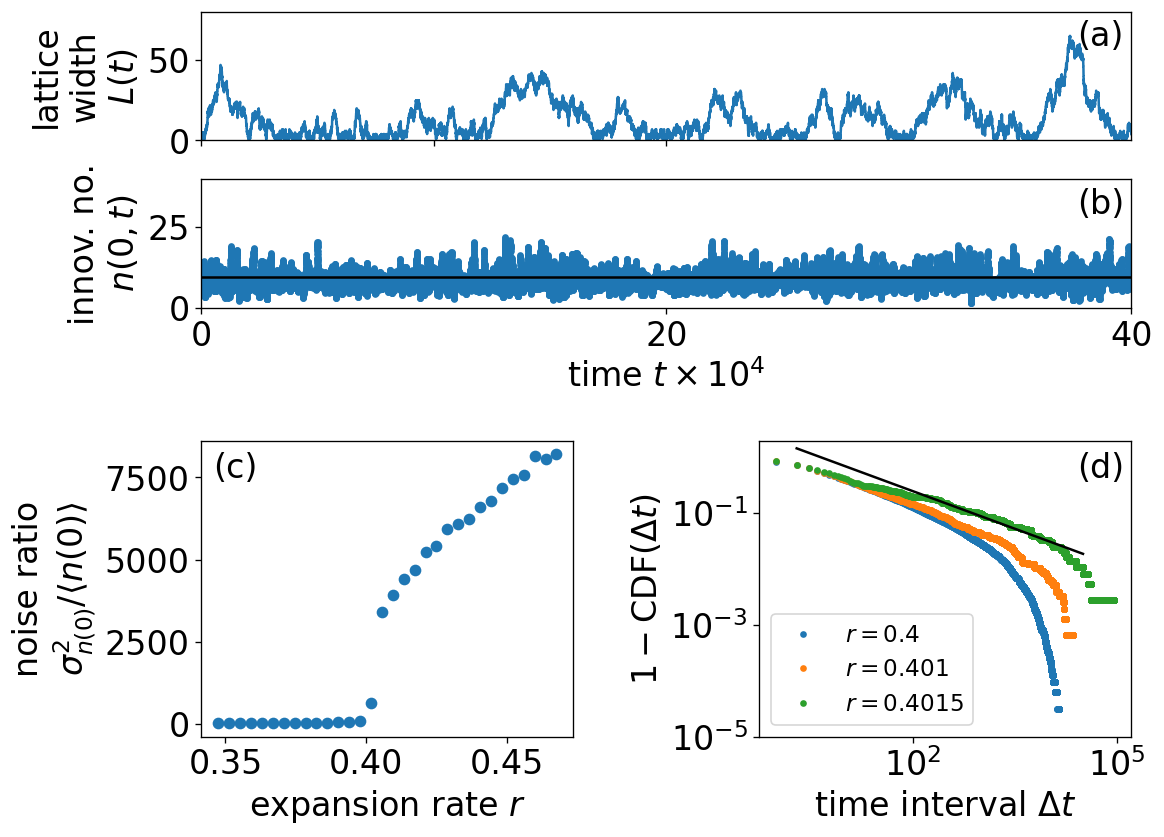}
	\caption{Large fluctuations at the critical point generate flourishing periods that eventually collapse. (a,b) Time series from automaton simulation near critical point $r=0.4$ for lattice width and innovation front number, respectively. (c) Normalized variance at innovation front around critical replication rate $r$. Critical point seems to be continuous because of slow, sublinear scaling of lattice size with simulation time beyond the critical point, but simulations show that lattice length will diverge for sufficiently long times. (d) Distribution of time periods of flourishing ideas defined as periods during which lattice width was greater than the mean.}\label{gr:critical re}
\end{figure}

Eq~\ref{si eq:corrected L} suggests that the dynamics of lattice width are determined by a competition between forces that drive the system towards 0 such as via small innovativeness $I\rightarrow 0$ and those that drive the system towards infinite growth. In a randomly growing system, we expect such gyrations to be most prominent when slight changes to the innovation front number $n(0,t)$ push us quickly from one extreme to another. Indeed, we noted exactly this in the collapsed limit in Eq~\ref{eq:obs rate}, where for $GIr\rightarrow0$, the width of the stable regime shrank to 0. Recalling that we had substituted in stationary values for $n(0)$, we recognize the essential role of the innovation front number $n(0)$ in determining the balance of the innovation and obsolescence fronts. This suggests as an order parameter, the typical fluctuation, or the noise-to-signal ratio, of the innovation front number,
\begin{align}
	\sigma^2 / \mu &= \left(\br{n(0,t)^2} - \br{n(0,t)}^2\right)/\br{n(0,t)}.\label{si eq:n0 order param}
\end{align}
We use Eq~\ref{si eq:n0 order param} characterize the fluctuations in the system when $I\ll1$, and we approach the critical point $r^*$ in Fig.~\ref{gr:critical re}. The appearance of a critical point is clear in the normalized variance of the order parameter $n_0$, which takes finite values beyond the critical $r$. As we approach the critical point, we see increasingly large fluctuations in the width of the lattice as shown in the example trajectory in Fig.~\ref{gr:critical re}A. The maximum durations for which the lattice breaks away from the collapsed state $\Delta t$ become longer as we approach the critical point, extending the cutoff of a heavy-tailed distribution as in Fig.~\ref{gr:critical re}C. Short of the critical point, the system eventually returns to the collapsed configuration.

%\begin{figure}\centering
%	\includegraphics[width=.8\linewidth]{mft_critical_points}
%	\caption{Comparison of innovation front number $n(0)$ and lattice length $L$ from mean-field approximations and automaton, or ``auto.'', model (averaged over independent trajectories). (b) Corrected approximation (Eq~\ref{si eq:corrected L}) predicts the absence of a critical point with respect to $r_o$.}\label{si gr:mft/auto critical}
%\end{figure}

\begin{figure}\centering
	\includegraphics[width=\linewidth]{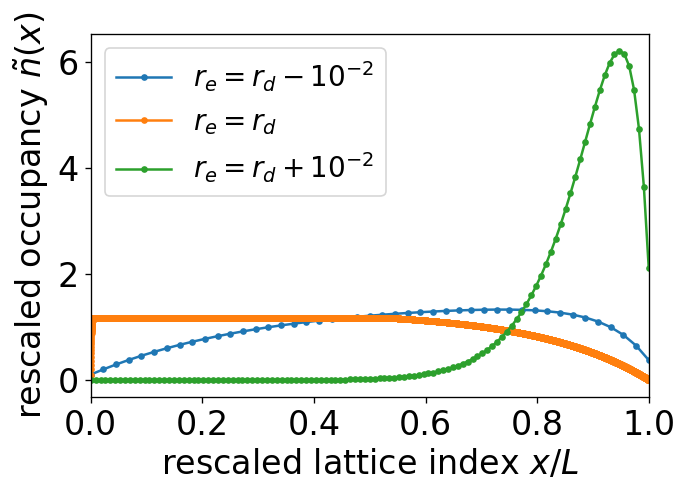}
	\caption{Examples of pseudogap types. Sublinear (blue), none (orange), and superlinear (green). These have been rescaled along both axes to facilitate comparison.}\label{si gr:pseudogap shapes}
\end{figure}

\begin{figure}\centering
	\includegraphics[width=\linewidth]{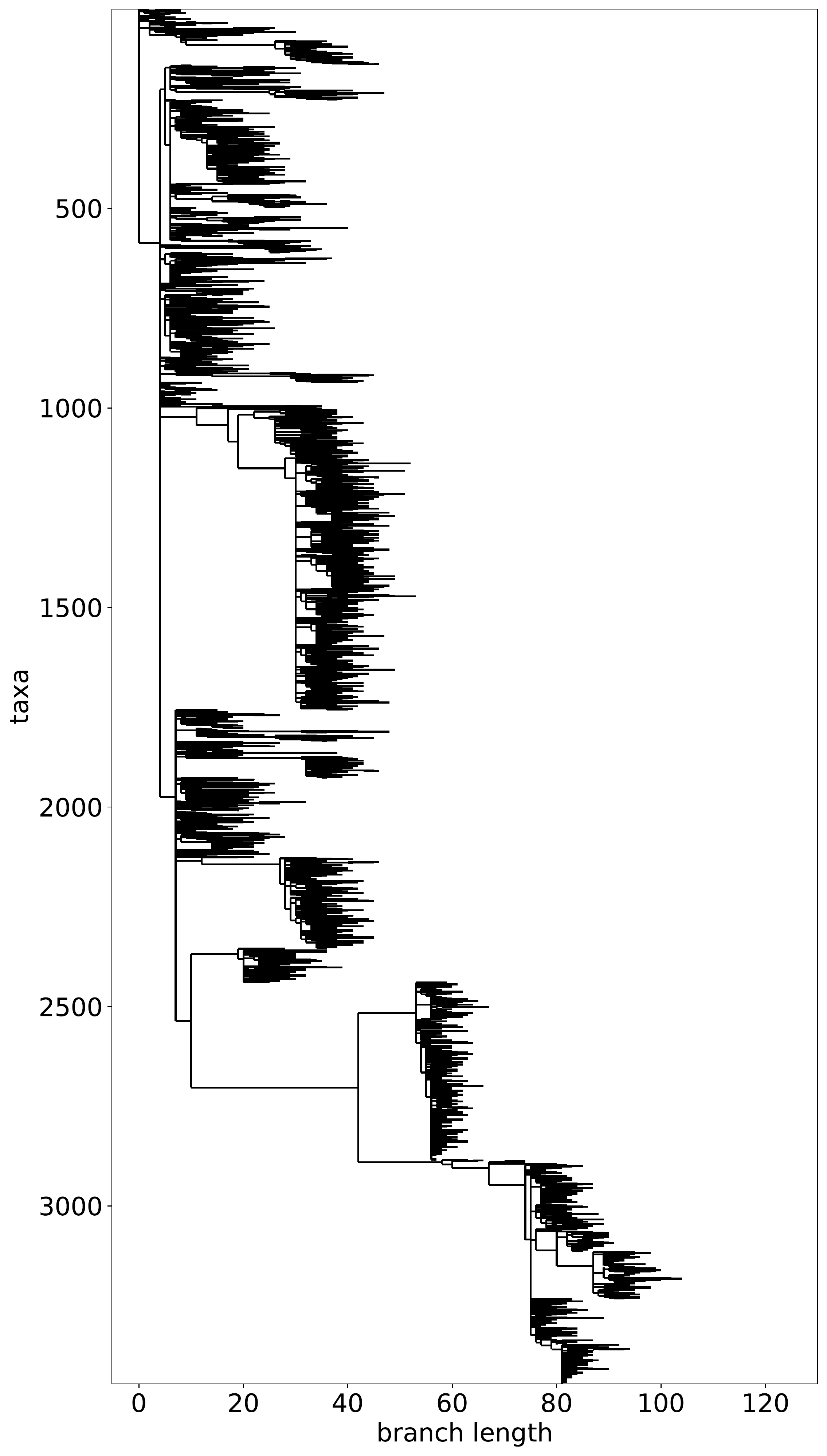}
	\caption{Example SARS-CoV-2 phylogeny tree from Nextstrain for North America.}
\end{figure}

\begin{figure}\centering
	\includegraphics[width=\linewidth]{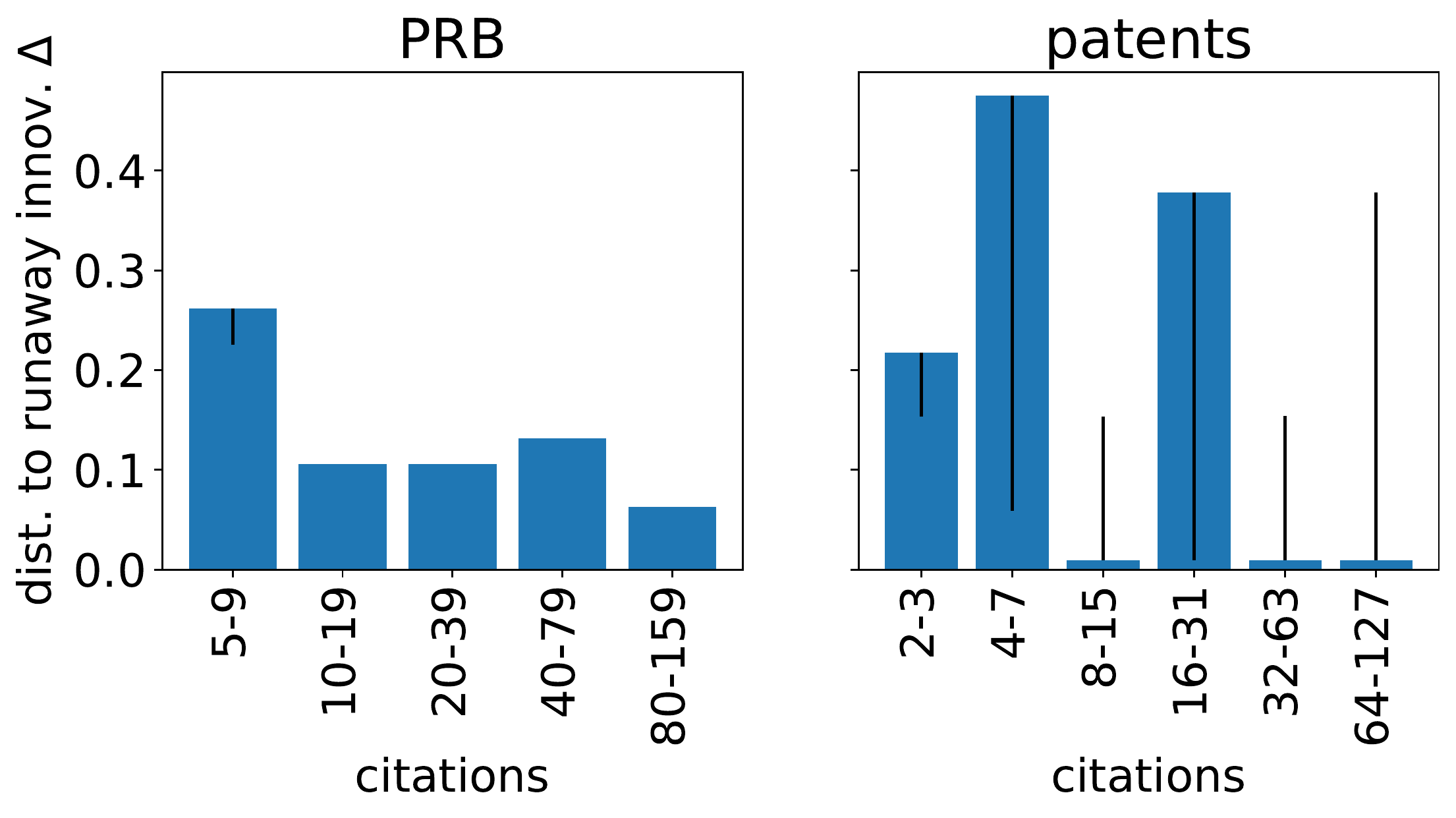}
	\caption{Distance to runaway innovation for all patent and {\it PRB} citation classes.}\label{si gr:cites}
\end{figure}

\end{document}